\documentclass{aa}  

\usepackage{hyperref}
\usepackage{graphicx}
\usepackage{txfonts}
\usepackage{xspace}   

\newcommand{\HI}{H\,\textsc{i}\xspace}
\newcommand{\Msol}{$M_{\odot}$\xspace}
\newcommand{\Mstar}{$M_\mathrm{\star}$\xspace}
\newcommand{\MHI}{$M_\mathrm{HI}$\xspace}
\newcommand{\NHI}{$N_\mathrm{HI}$\xspace}
\newcommand{\kms}{km\,s$^{-1}$\xspace}
\newcommand{\Mvir}{$M_\mathrm{vir}$\xspace}
\newcommand{\Rvir}{$R_\mathrm{vir}$\xspace}

\newcommand{\degree}{$^\circ$\xspace}
\newcommand{\SoFiA}{\texttt{SoFiA}\xspace}

\begin{document}

   \title{The MeerKAT Fornax Survey
 II. The rapid removal of \HI from dwarf galaxies in the Fornax cluster}

   \titlerunning{\HI in Fornax dwarfs}


   \author{D.~Kleiner\inst{1, 2}, P.~Serra\inst{2}, F.~M.~Maccagni\inst{1, 2}, M.~A.~Raj\inst{3}, W.~J.~G.~de Blok\inst{1, 3, 4}, G.~I.~G.~Józsa\inst{5, 6}, P.~Kamphuis\inst{7}, R.~Kraan-Korteweg\inst{4}, F.~Loi\inst{2}, A.~Loni\inst{8, 2}, S.~I.~Loubser\inst{9}, D.~Cs.~Molnár\inst{2}, T.~A.~Oosterloo\inst{1, 3}, R.~Peletier\inst{3} \and D.~J.~Pisano\inst{4}}
   \authorrunning{Kleiner et al.}

  \institute{Netherlands Institute for Radio Astronomy (ASTRON), Oude Hoogeveensedijk 4, 7991 PD Dwingeloo, the Netherlands\\
              \email{kleiner@astron.nl}
              \and
              INAF - Osservatorio Astronomico di Cagliari, Via della Scienza 5, I-09047 Selargius (CA), Italy
              \and
              Kapteyn Astronomical Institute, University of Groningen, PO Box 800, NL-9700 AV Groningen, the Netherlands
              \and
              Department of Astronomy, University of Cape Town, Private Bag X3, Rondebosch 7701, South Africa
              \and
              Max-Plank-Institut f\"ur Radioastronomie, Auf dem H\"ugel 69, 53121
              \and
              Department of Physics and Electronics, Rhodes University, PO Box 94, Makhanda, 6140, South Africa
              \and
              Ruhr University Bochum, Faculty of Physics and Astronomy, Astronomical Institute, 44780 Bochum, Germany
              \and
              Armagh Observatory and Planetarium, College Hill, Armagh BT61 9DG, UK
              \and
              Centre for Space Research, North-West University, Potchefstroom 2520, South Africa \\}

   \date{Received 20 March 2023 / Accepted 18 May 2023}
    

  \abstract{We present MeerKAT Fornax Survey atomic hydrogen (\HI) observations of the dwarf galaxies located in the central $\sim2.5 \times 4$ deg$^2$ of the Fornax galaxy cluster (\Rvir $\sim2^\circ$). The \HI images presented in this work have a $3\sigma$ column density sensitivity between 2.7 and $50\times10^{18}$ cm$^{-2}$ over 25\,\kms for spatial resolution between 4 and 1\,kpc. We are able to detect an impressive \MHI = 5 $\times$ 10$^{5}$ \Msol 3$\sigma$ point source with a line width of 50\,\kms at a distance of 20\,Mpc. We detected \HI in 17 out of the 304 dwarfs in our field, with 14 out of the 36 late-type dwarfs (LTDs) and three out of the 268 early-type dwarfs (ETDs). The \HI-detected LTDs have likely just joined the cluster and are on their first infall as they are located at large clustocentric radii, with comparable \MHI and mean stellar surface brightness at fixed luminosity as blue, star-forming LTDs in the field. By contrast, the \HI-detected ETDs have likely been in the cluster longer than the LTDs and acquired their \HI through a recent merger or accretion from nearby \HI. Eight of the \HI-detected LTDs host irregular or asymmetric \HI emission and disturbed or lopsided stellar emission. There are two clear cases of ram pressure shaping the \HI, with the LTDs displaying compressed \HI on the side closest to the cluster centre and a one-sided, starless tail pointing away from the cluster centre. The \HI-detected dwarfs avoid the most massive potentials (i.e. cluster centre and massive galaxies), consistent with massive galaxies playing an active role in the removal of \HI. We created a simple toy model to quantify the timescale of \HI stripping in the cluster by reproducing the observed $M_{r'}$ - \MHI relation. We find that a \MHI = 10$^{8}$ \Msol dwarf is stripped in $\sim$ 240\,Myr. The model is consistent with our observations, where low-mass LTDs are directly stripped of their \HI from a single encounter and more massive LTDs can harbour a disturbed \HI morphology due to longer times or multiple encounters being required to fully strip their \HI. This is the first time dwarf galaxies with \MHI $\lesssim$ 1 $\times$ 10$^{6}$ \Msol have been detected and resolved beyond the local group and in a galaxy cluster.}
  


   \maketitle
%
\section{Introduction}
Dwarf galaxies are the most abundant type of galaxy throughout the Universe. In general, dwarf galaxies have weak gravity, a metal poor interstellar medium (ISM), low interstellar pressure, insufficient dust shielding, and the disks tend to be thick and diffuse \citep[see][for a recent review]{Henkel2022}. Due to these properties, dwarf galaxies are the most sensitive to internal processes, such as star formation and feedback, and non-secular processes, such as tidal and hydrodynamical interactions that strip the ISM and distort the stellar body. This makes them the best candidates to test and understand the rapid and efficient processes that evolve star-forming late-type dwarfs (LTDs) into quiescent early-type dwarfs (ETDs). It is well established that the dense environment of galaxy clusters provide hostile conditions that efficiently produce quiescent galaxies of any mass via ISM removal and the quenching of star formation \citep[for a comprehensive review, see e.g.][]{Boselli2006, Boselli2014c, Cortese2021, Boselli2022}. Direct stripping of the cold ISM either through hydrodynamical, tidal, or a combination of both forces is required during cluster infall to explain the lack of ISM observed in cluster galaxies \citep[e.g.][]{Valluri1990, Valluri1991, Mayer2006, deRijcke2010, Gavazzi2018, Boselli2022}. Both observations and simulations show that low-mass galaxies falling into clusters have their ISM fully stripped close to or just past their first pericentric passage, resulting in a more rapid quenching \citep[e.g.][]{Boselli2008, Boselli2014a, Pasquali2019, Lotz2019}.

Atomic neutral hydrogen (\HI) remains one of the best observational tracers for measuring environmentally driven galaxy transformations \citep[e.g.][]{Cortese2010, Cortese2011, Catinella2018, deBlok2018}. However, dwarf galaxies in all environments are of a low mass, have small angular sizes, and are often diffuse in nature. Therefore, being able to simultaneously detect and resolve \HI in these galaxies has remained laborious for many years. This challenge is even more acute inside galaxy clusters given the aforementioned processes of ISM removal. Despite these observational constraints, there are multiple studies in the literature that have detected \HI in dwarf galaxies in a variety of environments. Surveys such as the Faint Irregular Galaxies GMRT Survey \citep[FIGGS;][]{Begum2008}, the Local Irregulars That Trace Luminosity Extremes - The \HI Nearby Galaxy Survey \citep[LITTLE-THINGS;][]{Hunter2012}, and the Local Volume \HI Survey \citep[LVHIS;][]{Koribalski2018} have detected \HI in star-forming LTDs in the field and groups down to \MHI $\sim$ 10$^{7}$ \Msol within 10\,Mpc. Similarly, \citet{Kovac2009} conducted a blind \HI survey of the Canes Venatici (CVn) volume ($\leq$ 18\,Mpc) that detected \HI in 70 star-forming LTDs, with a \HI mass range of 10$^{6.5}$--10$^{9.8}$ \Msol. The observations of nearby LTDs have shown that scaling relations such as the \HI-to-optical-diameter ratio hold true in this mass range and that LTDs in low-density environments can host asymmetric and irregular \HI morphologies. This is further supported by the irregular morphology and asymmetric velocity fields in the outer \HI disk of 18 starbursting LTDs that was likely caused by recent interactions \citep{Lelli2014a}. In the Virgo galaxy cluster, $\sim$80 dwarfs between \MHI = 10$^{6.9}$--10$^{7.7}$ \Msol (with 25\% completeness at the low-mass end) were detected in \HI by the ALFALFA survey \citep{Giovanelli2005, Huang2012a, Hallenbeck2012}. The majority were LTDs; although, a few were ETDs akin to transitional galaxies, which have little ongoing star formation despite the detectable \HI reservoir \citep{Hallenback2017}. Recently, \citet{Poulain2022} used the ALFALFA and ATLAS$^{\rm 3D}$ \HI \citep{Cappellari2011, Serra2012a} surveys to detect \HI in 145 dwarf satellite galaxies observed in the Mass Assembly of early-Type GaLAxies with their fine Structures (MATLAS) deep imaging survey and found that $\sim$ 80\% of dwarf satellites increase their \HI mass as a function of projected distance from the central galaxy. While progress has been made in understanding the nature of \HI in dwarf galaxies, the aforementioned work has been limited to the local volume with restrictions on the combined \HI sensitivity and resolution. 

If dwarf galaxies are observed in the dense cluster environment, we are able to constrain galaxy quenching mechanisms on galaxies that are transformed and quenched the quickest. The Fornax cluster is a small, low-mass \citep[\Mvir = 5 $\times$ 10$^{13}$ \Msol;][]{Drinkwater2001} cluster that is located 20\,Mpc away \citep{Blakeslee2009}, making it the second closest galaxy cluster to Earth after Virgo. Fornax is more dynamically evolved than Virgo; although, it is not virialized as there are massive star-forming late types present as well as in-falling groups \citep[e.g NGC\,1365 and NGC\,1316][]{Drinkwater2001, Raj2019, Kleiner2021, Loni2021}. The intracluster medium (ICM) of Fornax is half the density; although, the galaxy density is twice as high in comparison to Virgo \citep[e.g.][]{Jordan2007}, implying that in-falling galaxies experience a lower ram-pressure force but are prone to tidal interactions with cluster galaxies. The galaxy population of Fornax is dominated by red, quiescent ETDs, implying that the cluster is driving the transformation of LTDs into ETDs \citep{Drinkwater2001, Schroder2001, Waugh2002, Venhola2018, Loni2021}.

In this paper we present a multi-wavelength analysis of the dwarf galaxies located in the central $\sim$$2.5 \times 4$ deg$^2$ of the Fornax galaxy cluster (\Rvir $\sim$$2$\degree), focussing on the \HI content that has been observed in the MeerKAT Fornax Survey \citep[MFS:][]{Serra2023}. The MeerKAT telescope \citep{Camilo2018} has exquisite column density sensitivity and resolution. For this work, we detected and resolved dwarf galaxies for the first time with \MHI $\lesssim$ 10$^{6}$ \Msol at a distance of 20\,Mpc. Previously, that had only been possible for Milky-Way satellites and dwarf galaxies in the local group. We made use of the deep optical imaging from the Fornax Deep Survey \citep[FDS:][]{Iodice2016, Venhola2018, Peletier2020} to use the morphological and photometric properties of the dwarfs as well as their spatial distribution in the cluster. Our main goals are to determine which dwarfs have \HI, which dwarfs are being transformed in the cluster, how they are being transformed, and on what timescale. 

This paper is organised as follows: Section \ref{sec:obs} describes the MFS observations and data reduction used in this work and the relevant optical images and catalogues. In Sect. \ref{sec:sample}, we define our dwarf galaxy sample. We present our main results in Sect. \ref{sec:results}, where we display the \HI images and show how the \HI relates to the optical colour, luminosity, and mean surface brightness, along with the location of the \HI detection relative to the cluster substructure and distance to the nearest massive galaxy. In Sect. \ref{sec:discussion} we discuss the mechanisms that remove \HI from dwarf galaxies in the cluster and in Sect. \ref{sec:toymodel} we make a simple toy model to measure the timescale of \HI removal. Finally, we summarise our results in Section \ref{sec:conclusion}. Throughout this paper, we assume a luminosity distance of 20\,Mpc to the Fornax cluster and all associated galaxies \citep{Blakeslee2009}. At this distance, 1\arcsec\xspace corresponds to 97\,pc.

\section{Observations}
\label{sec:obs}
\subsection{MeerKAT radio data}

With reference to Table 2 of \citet{Serra2023}, in this work we measure \HI parameters (e.g. flux, line width) from the MFS data products\footnote{Available at the MFS website -- \url{https://sites.google.com/inaf.it/meerkatfornaxsurvey}} at a resolution of $66$\arcsec, that has a 3$\sigma$ column density sensitivity of 1.3 $\times$ 10$^{18}$\,cm$^{-2}$ over 25\,\kms, and is the most likely to recover all of the \HI flux. Given the availability of multiple resolutions sensitive to different \HI column densities, we present the \HI images with a resolution of $11$\arcsec, $21$\arcsec\xspace and $41$\arcsec, which cover a 3$\sigma$ column density sensitivity range of 2.7--50 $\times$ 10$^{18}$\,cm$^{-2}$ over 25\,\kms for spatial resolutions between 4 and 1\,kpc. The $41$\arcsec\xspace cube has the lowest rms = 0.24\,mJy\,beam$^{-1}$, that enables us to detect a 3$\sigma$ point source with a line width of 50\,\kms down to a \MHI = 5 $\times$ 10$^{5}$ \Msol at the distance of the Fornax cluster.

We limit our analysis to the portion of the MFS that has been completed at the time of writing. To avoid unreliable \HI parameterisation and poor image fidelity due to the rapidly changing rms at the mosaic edge, we only consider sources where the noise is at less than twice the minimum rms. The resulting field is $\sim$$2.5 \times 4$ deg$^2$ with our most sensitive 3$\sigma$ \HI mass limit of \MHI = 5 $\times$ 10$^{5}$\,\Msol and least sensitive 3$\sigma$ \HI mass limit of \MHI = 1 $\times$ 10$^{6}$\,\Msol for a 50\,\kms point source.

\subsection{Optical data}
We use deep optical images from the FDS \citep{Iodice2016, Venhola2018}, which are excellent for detecting dwarfs and resolving low surface brightness galaxies. The FDS observed the Fornax cluster using the OmegaCAM wide-field imager on the VLT Survey Telescope (VST). Observations were taken in the Sloan Digital Sky Survey (SDSS) $u^{\prime}$, $g^{\prime}$, $r^{\prime}$, and $i^{\prime}$-bands and are sensitive down to an impressive surface brightness limit of 28.4 mag arcsec$^{2}$ in the $g^{\prime}$-band. We use the photometric catalogue from \citet{Su2021}, who used a combination of aperture photometry, Sérsic + point spread function (PSF) and multi-component decompositions to quantify the light distribution of each massive cluster galaxy \citep{Iodice2019, Raj2019, Raj2020} and the dwarf galaxies \citep{Venhola2018}. 

One caveat of the FDS is that the VST is unable to image the Fornax cluster where extremely bright foreground stars are present. There are two dwarfs we detect in \HI (ESO\,358-G064 and FCC\,323) located in a region of the cluster where FDS images are either unavailable or not observed in all bands. To rectify this, we use images and the photometric catalogue from the Dark Energy Camera Legacy Survey (DECaLS) DR9\footnote{\url{https://www.legacysurvey.org/}} \citep{Dey2019}, in the same SDSS filters. We adopt the same method as \citet{Su2021} and use the photometric-colour relation  to estimate the stellar mass \citep{Taylor2011}, and aperture photometry to measure the projected surface brightness of ESO\,358-G064 and FCC\,323. The surface brightness limit of the DECaLS $g^{\prime}$-band images in this area is $\sim$27.9 mag arcsec$^{2}$, slightly shallower than the FDS does not present any issue for this work.

\section{Dwarf galaxy sample definition}
\label{sec:sample}
There are a variety of definitions in the literature of what constitutes a dwarf galaxy, and a common method is to use a simple magnitude cut \citep[e.g.][]{Hallenbeck2012} in a given volume. \citet{Venhola2018} created a catalogue of Fornax cluster dwarf galaxies in the FDS, using the definition of M$_{r^{\prime}}$ $>$ $-$18.5 with a minimum size of 2\arcsec, and separated the cluster dwarfs from background galaxies using the colour–magnitude, luminosity–radius and luminosity–concentration relations. We use this catalogue as the starting point in this work. However, the full catalogue contains some galaxies that are arguably not dwarfs. The best example is NGC\,1427A, which may indeed appear as a dwarf given its irregular morphology, although it has a stellar and \HI mass of $\sim$ 2 $\times$ 10$^{9}$ \Msol \citep{Lee-Waddell2018, Loni2021}, making it quite massive for a dwarf. The \citet{Venhola2018} catalogue used a combination of visual and parametric classifications to define the first order morphology of the Fornax cluster dwarfs.  

\begin{figure}

    \includegraphics[width = \columnwidth]{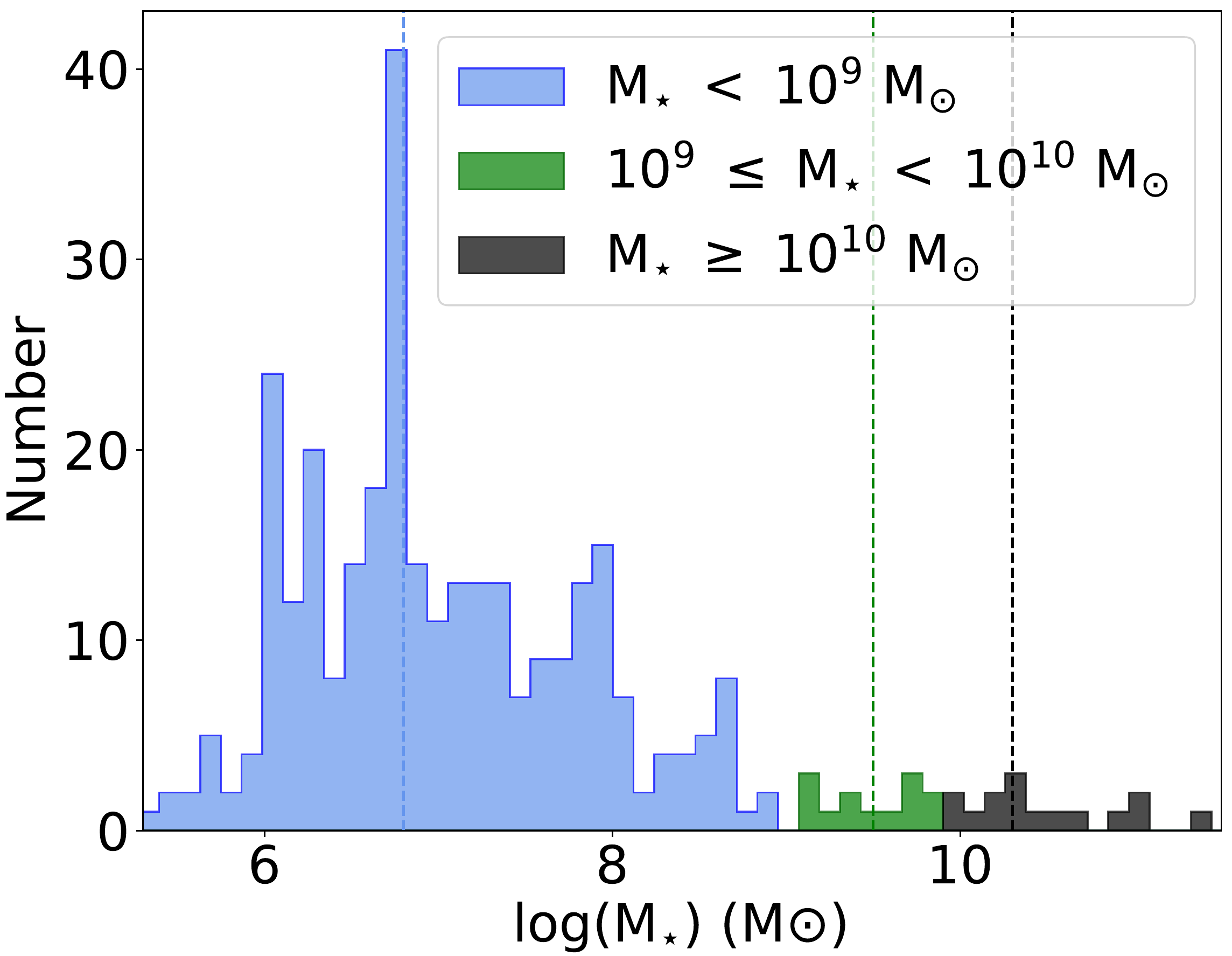}

\caption{Stellar mass distribution for our different galaxy samples. Each sample is defined by their stellar mass estimates \citep{Su2021}. The dwarfs are shown in light blue, the intermediate galaxies are in green and the massive galaxies are in black. The vertical dashed lines for each respective colour show the median of the samples.}

\label{fig:sample_mstellar}
\end{figure}

It has been suggested that at \Mstar $\approx$ 10$^{9}$ \Msol, galaxies transition from dwarfs to giants due to real structural differences \citep{Watkins2023}. In this work we are interested in the low-mass galaxy population that is the most efficiently transformed by the cluster environment. Therefore, using the stellar masses measured by \citet{Su2021}, we define dwarf galaxies as \Mstar $<$ 10$^{9}$ \Msol. We also define intermediate galaxies to be 10$^{9}$ $\leq$ \Mstar $<$ 10$^{10}$ \Msol which are intended to represent the populations of galaxies that are neither obvious dwarfs nor obvious giant galaxies. Lastly, we define the massive galaxies those with \Mstar $\geq$ 10$^{10}$ \Msol. There are 304 dwarfs (36 LTDs and 268 ETDs), 13 intermediate mass and 15 massive galaxies within our field. We show the stellar mass distribution for each sample in Fig. \ref{fig:sample_mstellar} and their basic sample statistics in table \ref{tab:samples}.  

\begin{table}
  \begin{center}
    \caption{Number of galaxies, median, and standard deviation of the stellar mass for our galaxy samples. Readers can refer to Fig. \ref{fig:sample_mstellar} for their stellar mass distribution.}
  \label{tab:samples}
    \begin{tabular}{c c c c}
        \hline
        \hline
        Sample & Ngals & Median M$\star$ & Std. M$\star$ \\
         &  & (\Msol) & (\Msol) \\
        \hline        
        Dwarf & 304 & 6.3 $\times$ 10$^{6}$ & 1.1 $\times$ 10$^{8}$ \\
        Intermediate & 13 & 3.2 $\times$ 10$^{9}$ & 1.9 $\times$ 10$^{9}$ \\ 
        Massive & 15 & 2.0 $\times$ 10$^{10}$ & 6.1 $\times$ 10$^{10}$ \\
        \hline
    \end{tabular}
  \end{center}

\end{table}

\section{\HI in cluster dwarfs}
\label{sec:results}
\begin{figure*}

    \centering
    \includegraphics[width = \textwidth]{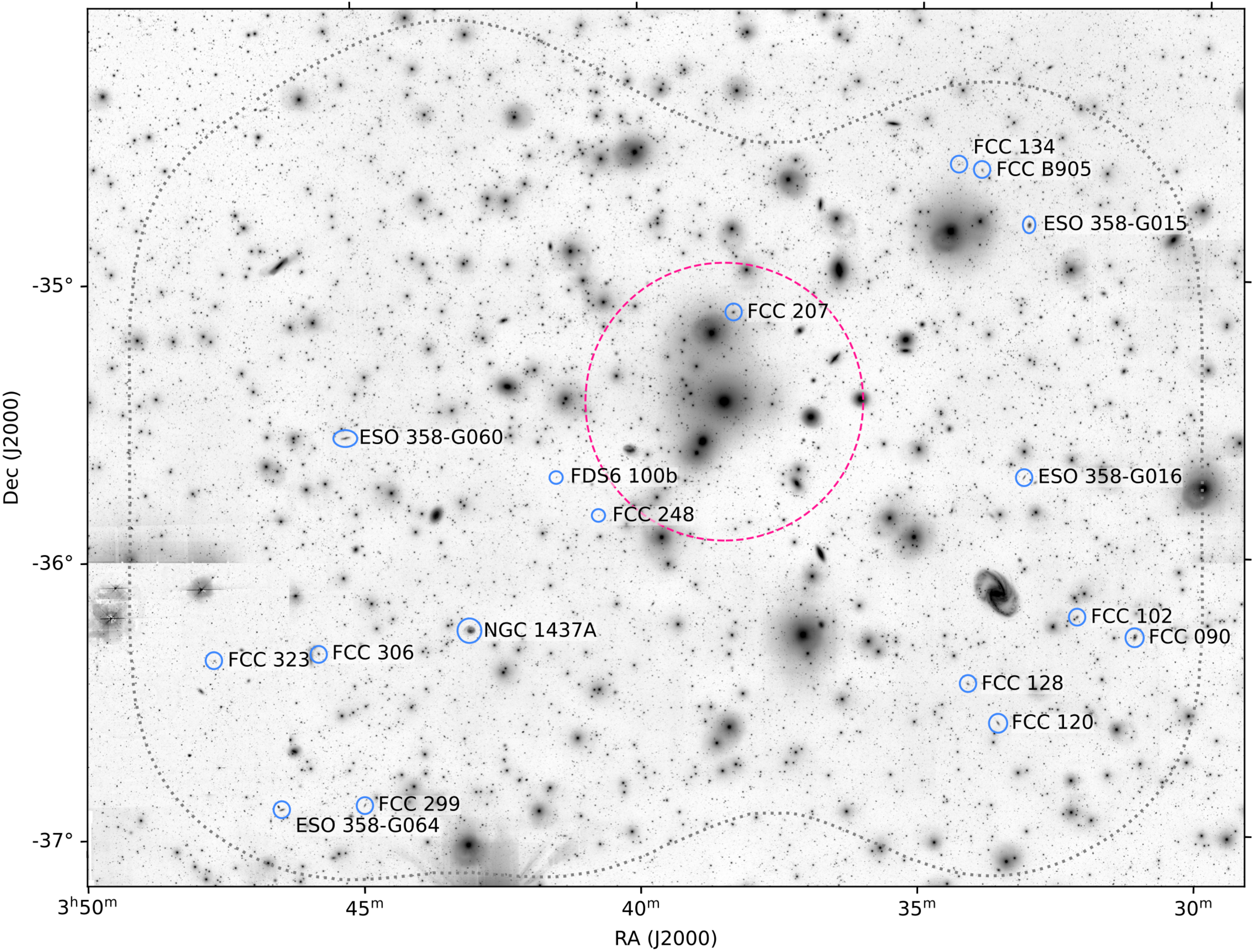}

\caption{\HI-detected dwarfs in the Fornax cluster. The background is the FDS $g^{\prime}$-band optical image, except in the region of FCC\,323 that has been filled in with DECaLS images. The dwarfs detected in \HI are shown in the blue ellipses with their respective labels. The dotted grey border denotes the $\sim2.5 \times 4$ deg$^2$ area analysed in this work. The pink dashed circle shows 0.25\Rvir.}

\label{fig:dwarf_locations}
\end{figure*}

\begin{figure*}

    \includegraphics[width = \textwidth]{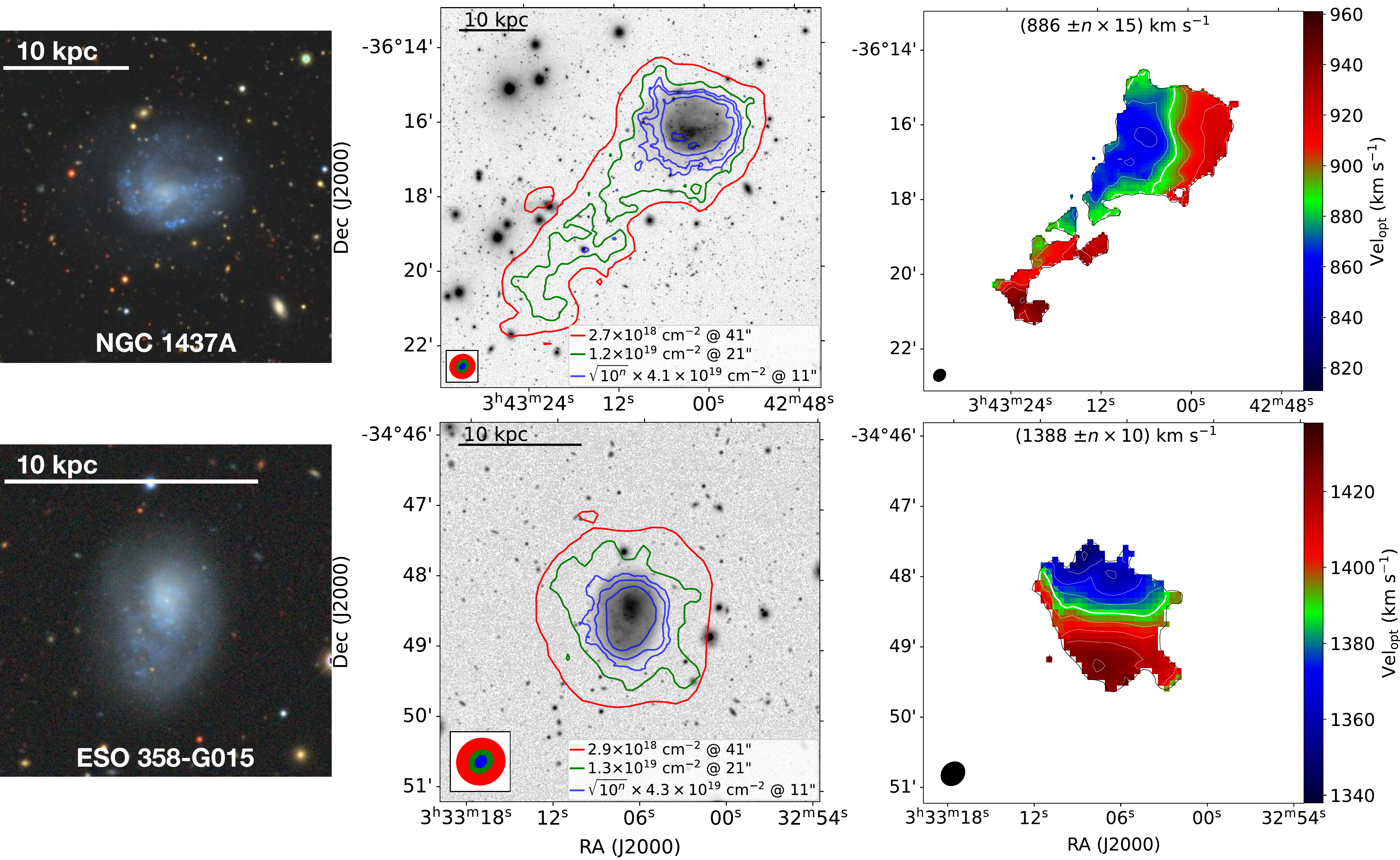}

\caption{Preview of the dwarf optical cutouts, multi-resolution MFS \HI overlays and velocity fields. The full figure with all the \HI detections are shown in Fig. \ref{fig:overlays}. The left panel shows the optical three-colour postage stamp of each dwarf, the middle panel shows the \HI contours overlaid on the FDS (DECaLS for FCC\,323) optical image where the blue contours show the $11$\arcsec\xspace \HI emission above 3$\sigma$ in steps of $\sqrt{10^n}$ and the green and red contours show the 3$\sigma$ \HI emission from the $21$\arcsec\xspace and $41$\arcsec\xspace images. The lowest column density sensitivities are labelled in accordance with the local rms over a 25 \kms line-width and the synthesised beam for each resolution are shown in the bottom left inset with their respective colours. The right panel shows the $21$\arcsec\xspace \HI velocity field. The systemic velocity is the thick white contour with the value shown up the top and the synthesised beam is shown in black on the bottom left.}

\label{fig:overlays_prev}
\end{figure*}

We detect \HI in 17 out of the 304 dwarfs within our field and show their position in the cluster in Fig.\ref{fig:dwarf_locations}. The detections include 14 out of the 36 LTDs, and 3 out of the 268 ETDs. We present a preview in Fig. \ref{fig:overlays_prev} and all the dwarfs detected in \HI in Fig. \ref{fig:overlays} that shows the DECaLS 3-colour postage stamp, the FDS optical image overlaid with \HI contours from the $11$\arcsec, $21$\arcsec\xspace and $41$\arcsec\xspace MFS images and the 21\arcsec\xspace velocity field. The basic \HI properties for the \HI detected dwarf galaxies are presented in Table \ref{tab:HI_prop} and the optical morphology, \HI mass, stellar mass, photometry and projected distance to nearest massive galaxy in Table \ref{tab:gen_prop}. Eight of the LTDs with \HI show clear signs of irregular or asymmetric \HI emission. We denote these galaxies as being (\HI) disturbed as the irregularities can be seen in the form of \HI tails, significant warps or non-rotating components (Fig. \ref{fig:overlays}) making them the most likely LTDs currently being shaped by the cluster environment. It is possible that more LTDs are disturbed; however, their signs are subtler. We refrain from labelling the ETDs as disturbed as \HI-bearing ETDs are an unusual class of galaxy at a different evolutionary stage compared to LTDs. 

\begin{table*}
  \begin{center}
    \caption{\HI properties of dwarfs measured from the $66$\arcsec MFS \HI cube. The position, integrated flux, and $w_\mathrm{20}$ are from the \SoFiA catalogue, while $v_\mathrm{sys}$ is estimated from where $11$\arcsec\xspace \HI iso-velocity intercepts the optical centre. The uncertainty in the integrated flux is a combination of the statistical error and 10\% uncertainty in the calibration of the flux scale \citep{Serra2023}.}
    \label{tab:HI_prop}
    \begin{tabular}{c c c c c c c}
        \hline
        \hline
        Name & Ra (J2000) & Dec (J2000) & Integrated flux  & $v_\mathrm{sys}$ & $w_\mathrm{20}$ & Disturbed\\
         & (hms) & (dms) & (Jy \kms) & (km\,s$^{-1}$) & (km\,s$^{-1}$) \\
        \hline    
        NGC\,1437A & 03:43:02.42 & -36:16:24.2 & 7.25 $\pm$ 0.7  & 886 & 128 & Y \\ 
        ESO\,358-G015 & 03:33:07.06 & -34:48:21.2 & 1.50 $\pm$ 0.2 & 1388 & 133 & Y \\
        ESO\,358-G016 & 03:33:09.19 & -35:43:05.2 & 1.26 $\pm$ 0.1 & 1700 & 131 & Y \\
        ESO\,358-G060 & 03:45:12.29 & -35:34:15.2 & 11.99 $\pm$ 1.2 & 803 & 129 & Y \\
        ESO\,358-G064 & 03:46:28.51 & -36:54:21.6 & 5.59 $\pm$ 0.7 & 1465 & 171 & N \\
        FCC\,090 & 03:31:08.28 & -36:17:24.4 & 0.59 $\pm$ 0.07 & 1825 & 129 & N \\
        FCC\,102 & 03:32:10.56 & -36:13:10.6 & 1.19 $\pm$ 0.1 & 1640 & 83 & Y \\
        FCC\,120 & 03:33:34.18 & -36:36:20.1 & 1.08 $\pm$ 0.1 & 846 & 60 & Y \\
        FCC\,128 & 03:34:07.03 & -36:27:57.6 & 0.14 $\pm$ 0.02 & 1543 & 68 & N \\
        FCC\,134 & 03:34:21.70 & -34:35:33.0 & 0.02 $\pm$ 0.01 & 1398 & 34 & N \\
        FCC\,207 & 03:38:19.25 & -35:07:44.8 & 0.0065 $\pm$ 0.006 & 1420 & 38 & N \\
        FCC\,248  & 03:40:43.37 & -35:51:38.9 & 0.010 $\pm$ 0.006 & 1859 & 15 & N \\ 
        FCC\,299 & 03:44:58.61  & -36:53:41.6 & 0.31 $\pm$ 0.04 & 2100 & 136 & N \\ 
        FCC\,306 & 03:45:45.38 & -36:20:46.0 & 2.03 $\pm$ 0.1  & 886 & 176 & Y \\
        FCC\,323 & 03:47:37.46 & -36:21:48.6 & 0.43 $\pm$ 0.05  & 1499 & 67 & N \\
        FCC\,B905 & 03:33:57.34 & -34:36:43.2 & 0.52 $\pm$ 0.06 & 1248 & 99 & Y \\
        FDS6\,100b & 03:41:27.84 & -35:43:13.1 & 0.14 $\pm$ 0.03 & 1983 & 50 & N \\ 
        \hline
    \end{tabular}
  \end{center}
\end{table*}

\begin{table*}
  \begin{center}
    \caption{Optical morphology, stellar mass, \HI mass, photometric properties, and projected distance to nearest massive galaxy for all the LTDs and the \HI-detected ETDs. The optical morphologies are taken from \citep{Venhola2018} and those with an asterisk are nucleated. The uncertainties in the \HI mass are propagated from the flux uncertainties in Table \ref{tab:HI_prop} and the upper limits are calculated as 3$\sigma$ of the local rms for a 50\,\kms wide point source. The stellar masses, $M_{r'}$ magnitudes, $g^{\prime} - r^{\prime}$ colours and $r^{\prime}$-band mean surface brightness out to one effective radius ($\mu_\mathrm{e}$) are taken from the \citet{Su2021} catalogue. The photometry for ESO\,358-G064 and FCC\,323 is taken from the DECaLS DR9 catalogue \citep{Dey2019}.} 
    \label{tab:gen_prop}
    \begin{tabular}{c c c c c c c c c}
        \hline
        \hline 
        Name & FCC number & Morphology & \HI mass & Stellar Mass & $R^{\prime}$ & $g^{\prime} - r^{\prime}$ & $\mu_\mathrm{e}$ & Dist. \\
         & & & (10$^{7}$ \Msol) & log(\Msol) & (mag) & (mag) & (mag\,arcsec$^{-2}$) & (kpc) \\
        \hline  
        NGC\,1437A & 285 & l & 68.4 $\pm$ 6.9 & 8.8 & -18.02 & 0.49 & 23.00 & 151 \\
        ESO\,358-G015 & 113 & l & 14.2 $\pm$ 1.4 & 8.5 & -17.03 & 0.5 & 22.87 & 181 \\
        ESO\,358-G016 & 115 & l & 11.9 $\pm$ 1.2 & 7.7 & -15.43 & 0.43 & 24.45 & 150 \\
        ESO\,358-G060 & 302 & l & 113.2 $\pm$ 11.3 & 7.8 & -16.02 & 0.15 & 24.92 & 148 \\
        ESO\,358-G064 & B1899 & l & 5.6 $\pm$ 0.7 & 8.5 & -16.20 & 0.59 & 23.13 & 415 \\
        FCC\,090 & 90 & l & 5.6 $\pm$ 0.6 & 8.9 & -17.60 & 0.56 & 21.23 & 180 \\
        FCC\,102 & 102 & l & 11.2 $\pm$ 1.2 & 7.8 & -15.80 & 0.47 & 23.47 & 103 \\
        FCC\,120 & 120 & l & 10.2 $\pm$ 1.0 & 7.9 & -15.72 & 0.43 & 24.28 & 161 \\    
        FCC\,128 & 128 & l & 1.3 $\pm$ 0.2 & 8.0 & -15.50 & 0.55 & 23.63 & 118 \\
        FCC\,134 & 134 & e & 0.2 $\pm$ 0.1 & 7.7 & -14.65 & 0.53 & 23.32 & 96 \\
        FCC\,207 & 207 & e & 0.06 $\pm$ 0.03 & 8.5 & -16.55 & 0.64 & 21.98 & 112 \\
        FCC\,248 & 248 & e & 0.1  $\pm$ 0.06 & 7.2 & -13.44 & 0.6 & 23.79 & 160 \\
        FCC\,299 & 299 & l* & 2.9 $\pm$ 0.4 & 7.8 & -15.18 & 0.5 &  23.83 & 372 \\
        FCC\,306 & 306 & l* & 19.1 $\pm$ 2.0 & 8.1 & -16.00 & 0.36 & 22.01 & 227 \\
        FCC\,323 & 323 & l & 4.0 $\pm$ 0.5 & 7.4 & -14.41 & 0.39 & 24.55 & 331 \\
        FCC\,B905 & B905 & l & 4.9 $\pm$ 0.6 & 7.8 & -15.12 & 0.49 & 23.69 & 125 \\
        FDS6\,100b & - & l & 1.3 $\pm$ 0.2 & 6.7 & -13.10 & 0.43 & 24.72 & 128 \\
        FCC\,247 & 247 & l & $\leq$ 0.06 & 7.2 & -14.11 & 0.31 & 24.54 & 131 \\
        FDS2\,393 & - & l & $\leq$ 0.07 & 6.3 & -11.34 & 0.36 & 24.88 & 117 \\
        FDS5\,010 & - & l & $\leq$ 0.06 & 5.7 & -9.41 & 0.69 & 26.15 & 119 \\
        FDS5\,137 & - & l & $\leq$ 0.06 & 6.6 & -12.31 & 0.48 & 24.07 & 155 \\
        FDS5\,159b & - & l & $\leq$ 0.06 & 5.7 & -9.64 & 0.56 & 26.17 & 223 \\
        FDS5\,608b & - & l & $\leq$ 0.10 & 5.7 & -10.34 & 0.5 & 25.71 & 395 \\ 
        FDS7\,021 & - & l & $\leq$ 0.08 & 6.3 & -11.72 & 0.61 & 24.54 & 411 \\
        FDS7\,085 & - & l & $\leq$ 0.06 & 6.0 & -10.99 & 0.28 & 24.66 & 317 \\
        FDS7\,275 & - & l & $\leq$ 0.06 & 6.1 & -11.54 & 0.19 & 24.31 & 223 \\
        FDS7\,461 & - & l & $\leq$ 0.06 & 6.3 & -11.34 & 0.54 & 26.36 & 174 \\
        FDS10\,130 & - & l & $\leq$ 0.08 & 6.1 & -11.13 & 0.46 & 25.30 & 118 \\
        FDS10\,167 & - & l & $\leq$ 0.09 & 5.5 & -9.97 & 0.61 & 27.13 & 77 \\ 
        FDS11\,403 & - & l & $\leq$ 0.06 & 6.4 & -12.07 & 0.27 & 26.15 & 102 \\
        FDS11\,411b & - & l & $\leq$ 0.06 & 6.2 & -11.42 & 0.56 & 24.38 & 106 \\ 
        FDS12\,052 & - & l & $\leq$ 0.09 & 6.8 & -12.98 & 0.4 & 23.84 & 287 \\
        FDS12\,135 & - & l & $\leq$ 0.07 & 5.6 & -10.12 & 0.33 & 25.64 & 274 \\
        FDS15\,085 & - & l & $\leq$ 0.06 & 5.8 & -10.03 & 0.25 & 26.54 & 58 \\
        FDS15\,148 & - & l & $\leq$ 0.07 & 6.1 & -11.02 & 0.38 & 24.76 & 122 \\
        FDS15\,155 & - & l & $\leq$ 0.06 & 6.0 & -11.25 & 0.26 & 24.92 & 184 \\
        FDS16\,141 & - & l & $\leq$ 0.06 & 6.0 & -11.21 & 0.4 & 25.04 & 105 \\
        FDS20\,100 & - & l & $\leq$ 0.11 & 6.9 & -13.24 & 0.29 & 23.98 & 285 \\
        FDS20\,502 & - & l & $\leq$ 0.09 & 5.9 & -10.25 & 0.5 & 24.71 & 162 \\
        \hline
    \end{tabular}
  \end{center}
\end{table*}

\subsection{New \HI dwarfs and features}
The following nine LTDs were detected in \citet{Loni2021}: NGC\,1437A, ESO\,358-G015, ESO\,358-G016, ESO\,358-G060, FCC\,090, FCC\,102, FCC\,120, FCC\,306 and FCC\,323. Therefore, this work has detected \HI for the first time in an additional eight dwarfs; five LTDs -- ESO\,358-G064, FCC\,128, FCC\,299, FCC\,B905, and FDS6\,100b; and three ETDs -- FCC\,134, FCC\,207, and FCC\,248. 

Due to the exquisite sensitivity and resolution of MeerKAT, we have detected and resolved new \HI features in some previously \HI-detected dwarfs. Here we describe the \HI features of the disturbed LTDs. NGC\,1437A and FCC\,306 have starless, \HI tails pointing away from the cluster centre, with velocity gradients aligned with the direction of the ram-pressure wind \citep{Serra2023}. In Fig. \ref{fig:overlays}, the $11$\arcsec\xspace image show a \HI compression into the edge of the stellar body on the side of the galaxy pointing towards the cluster centre, exactly where the ram-pressure force is the strongest. ESO\,358-G016 may also have a visible \HI compression in the $11$\arcsec\xspace image and potentially a \HI tail in the $21$\arcsec\xspace image on the opposite side of the compression. FCC\,102 shows a compression on the west in the $11$\arcsec\xspace image as well as a potential tail in the form of extended \HI emission to the east at all resolutions. FCC\,102 has a prominent S-shaped iso-velocity contour which is the discernible sign of a changing position angle (PA) due to a warp. While FCCB\,905 is not as well resolved as NGC\,1437A and FCC\,306, the $11$\arcsec\xspace image does suggest the \HI is being compressed on the side that faces the cluster centre and there may be a small \HI tail pointing in the opposite direction. ESO\,358-G015 has asymmetric \HI emission visible in the $21$\arcsec\xspace image and a U-shaped iso-velocity contour. ESO\,358-G060 has some clear asymmetries in the $11$\arcsec\xspace image with what appears to be an S-shaped iso-velocity contour containing kinks, also likely due to a warp. The well resolved velocity field of FCC\,120 does not show any clear rotation, which is unusual as it appears to be a blue, edge-on LTD.

The stellar emission is also clearly irregular for some of the \HI-disturbed LTDs (Fig. \ref{fig:overlays}). The following have been analysed by \citet{Raj2019}; NGC\,1437A, ESO\,358-G015, ESO\,358-G016, ESO\,358-G060 and FCC\,306. The stellar emission of NGC\,1437A is asymmetric \citep{Raj2019} and the DECaLS image (Fig. \ref{fig:overlays}) shows a diffuse stellar component in the north that may also be part of a halo. ESO\,358-G015 has a lopsided stellar tail pointing towards the cluster centre, ESO\,358-G016 appears to have a disturbed outer stellar disk, ESO\,358-G060 has a warped stellar disk with irregular star-forming regions and FCC\,306 is asymmetric with a disrupted outer stellar disk \citep{Raj2019}. The stellar emission for FCC\,B905 is clearly lopsided and this is also true, although to a lesser extent for FCC\,102. This is confirmed by the non-parametric morphological indices (e.g. Asymmetry, Clumpiness and Gini) measured by \citet{Su2021} who show that FCC\,102 is more asymmetric and has a higher Gini value compared to FCC\,120. 

We do not include FCC\,090 and FCC\,128 as disturbed LTDs, although they have some unusual features. The optical morphology of FCC\,090 was originally classified as a peculiar elliptical \citep{Ferguson1989} although it is a LTD in \citet{Venhola2018}. The FDS image shows some structure in the stellar body, such as a possible bar in the core. The \HI emission of FCC\,090 extends beyond the stellar body with hints of asymmetries and the velocity field does not appear to show rotation. There is a hint of a \HI tail for FCC\,128 shown in $21$\arcsec\xspace and $41$\arcsec\xspace images, and the velocity field suggests the presence of non-rotating components, although the \HI is not resolved well enough to state this definitively.

\subsection{The relation between \HI and stellar emission}
\begin{figure}

    \includegraphics[width = \columnwidth]{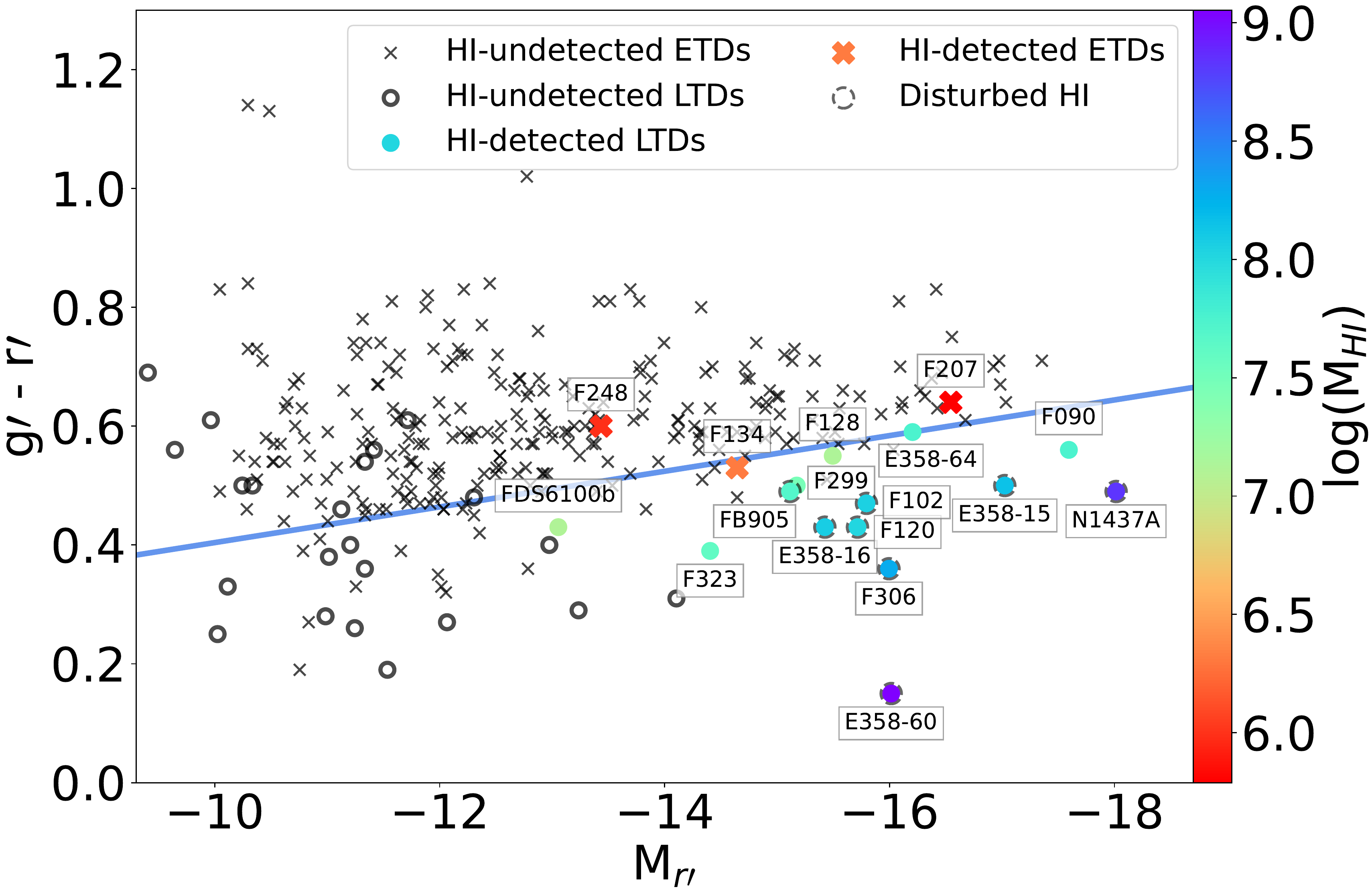}

\caption{Colour-magnitude diagram of dwarf galaxies within the $\sim$2.5 $\times$ 4 deg$^2$ area shown in Fig. \ref{fig:dwarf_locations}. Circles show the LTDs and crosses are the ETDs. The dwarfs detected in \HI have been labelled and coloured by their \MHI. The LTDs with \HI are shown as filled circles and the ETDs with \HI are the thick crosses. The LTDs with disturbed \HI morphologies are enclosed by dashed circles. The solid blue line separates the blue cloud from the red sequence. As expected, the bluer, brighter LTDs have a higher \MHI.}

\label{fig:CMR}
\end{figure}

In the following section, we discuss the galaxy properties listed in Table \ref{tab:gen_prop} and how they relate to each other. In Fig. \ref{fig:CMR}, we present the galaxy colour-magnitude diagram. We qualitatively separate the blue cloud from the red sequence by the relation $g^{\prime}$--$r^{\prime}$=$-$0.03$r^{\prime}$--0.1, that results in 10 out of the 36 LTDs in the red sequence (at the faint end) and 25 out of the 268 ETDs in the blue cloud. We detect all LTDs brighter than $M_{r'} = -14.2$ in \HI and as expected, \MHI increases for bluer, brighter LTDs. Two out of the three \HI-detected ETDs reside on the red sequence, although the third (FCC\,134) is significantly bluer and has the $\sim$ same colour as the LTDs FCC\,090 and ESO\,358-G064.

\begin{figure}

    \includegraphics[width = \columnwidth]{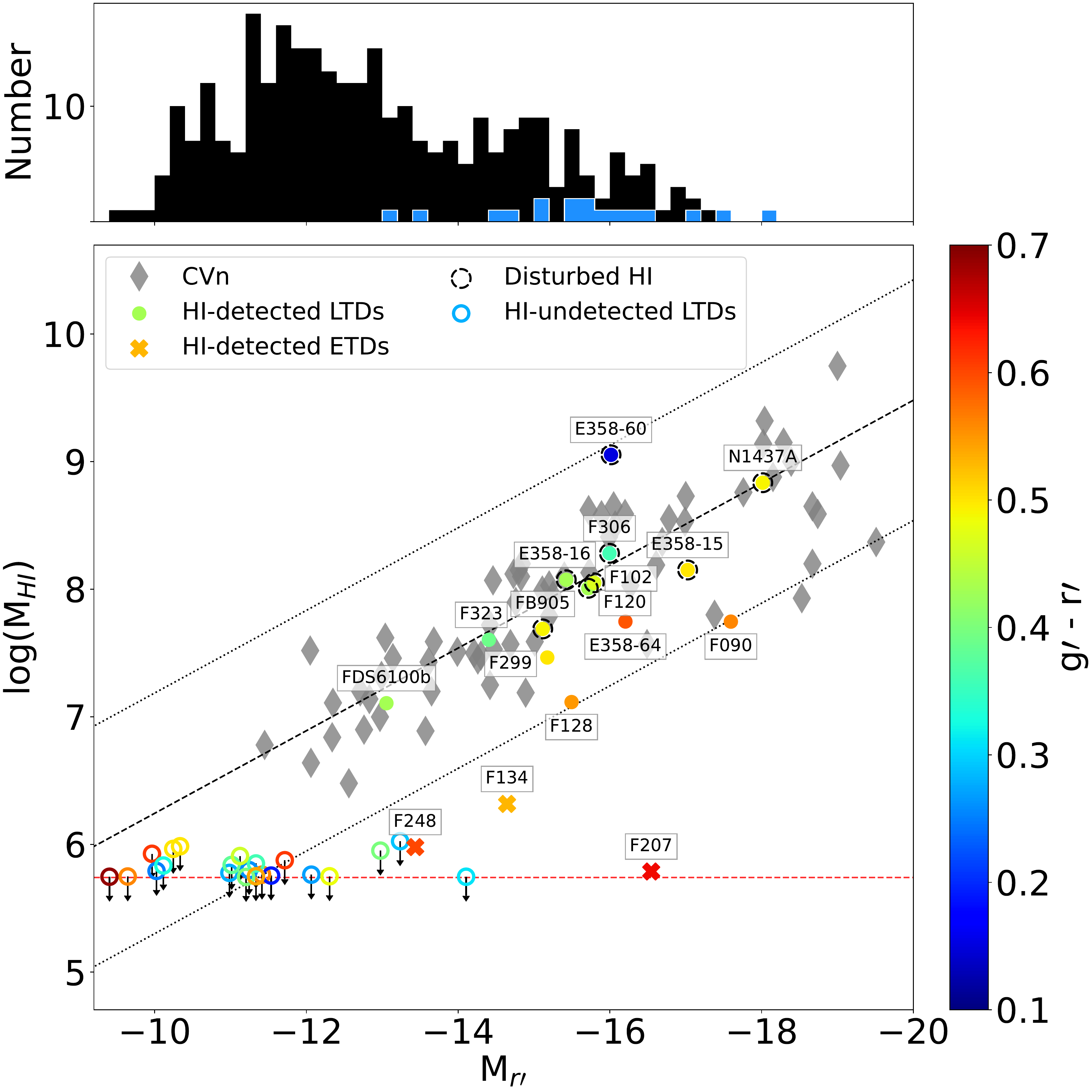}

\caption{Luminosity-\HI relation. The main panel shows $M_{r'}$ as a function of \MHI. A control sample of CVn dwarfs is shown in grey diamonds and the linear fit with 3$\sigma$ scatter in the range of -18 $<$ $M_{r'}$ $\leq$ -12 are shown as the black dashed and dotted lines, respectively. Fornax LTDs are shown as circles, where the filled circles have been detected in \HI and the open circles have not. Upper limits of the \HI-undetected LTDs are shown as downward arrows. Dashed circles enclose the LTDs with \HI-disturbed morphologies and thick crosses show the \HI-detected Fornax ETDs. All \HI-detected dwarfs are labelled and coloured by their $g^{\prime} - r^{\prime}$ colour. The dashed red line shows our \HI detection limit of \MHI = 5 $\times$ 10$^{5}$ \Msol. In the upper panel, the black histogram shows all dwarfs in our field, with \HI-detected dwarfs shown in blue. All of the \HI-detected LTDs lie within the 3$\sigma$ scatter of the CVn fit, implying that a significant amount of \HI has not been lost yet even those with disturbed \HI morphologies.}

\label{fig:lum_HI}
\end{figure}

We show the luminosity-\HI relation in Fig. \ref{fig:lum_HI} and compare our LTDs and \HI-detected ETDs to the CVn dwarf galaxies \citep{Kovac2007, Kovac2009}. Previous scaling relations have shown that \MHI is correlated with luminosity for late types brighter than $M_{r'} = -17$ \citep[e.g.][]{Denes2014}. The CVn dwarfs extends this trend down to fainter magnitudes of $M_{r'} = -12$. We quantify the luminosity-\HI relation of the CVn dwarfs with a linear fit in the range of -18$<$$M_{r'}$$\leq$-12 as our brightest \HI-detected LTD has a $M_{r'} = -18$. The dashed and dotted lines shows our linear fit and the 3$\sigma$ scatter, respectively (Fig. \ref{fig:lum_HI}). In the luminosity-\HI relation, we show the $g^{\prime} - r^{\prime}$ colours of our LTDs and \HI-detected ETDs. The \HI detections show that \HI-deficiency is correlated with redder colours. All our HI-detected LTDs are within 3$\sigma$ of the CVn fit and are therefore comparable to LTDs in the field. The \HI-disturbed LTDs are all brighter than $M_{r'} = -15$, have blue colours and mostly reside $\sim$ on the CVn fit, implying that a significant amount of \HI has not been lost yet. Irrespective of \HI morphology, if a LTD has \HI, it is within the scatter of the CVn fit, that is we do not detect any LTDs in \HI below the $M_{r'}$--\MHI relation. There are five \HI-undetected, blue LTDs between $M_{r'} = -12$ and $-14.2$ that are $>$3.5$\sigma$ from the CVn fit. Given their morphology, luminosity and colour, we would expect to detect these LTDs in \HI. The most extreme case is FCC\,247, which is 5.8$\sigma$ below the CVn fit. The upper limits for the \HI-undetected LTDs were calculated by using 3$\sigma$ of the local rms for a 50\,\kms wide point source. The \HI-undetected LTDs fainter than $M_{r'} = -12$ are within $\sim$3$\sigma$ of the extrapolated CVn fit and are $\sim$50\% split between red and blue colours. The \HI-detected ETDs are well below 3$\sigma$ of the CVn fit, although this is unsurprising given that the gas-fraction of ETDs and LTDs are significantly different, independent of luminosity. The histogram in the upper panel of Fig. \ref{fig:lum_HI} shows all of the dwarfs (black) and the \HI-detected dwarfs (blue) as a function of luminosity in our field and show that there are many dwarfs (mainly ETDs) that span the full luminosity range and are not detected in \HI.

\begin{figure}

    \includegraphics[width = \columnwidth]{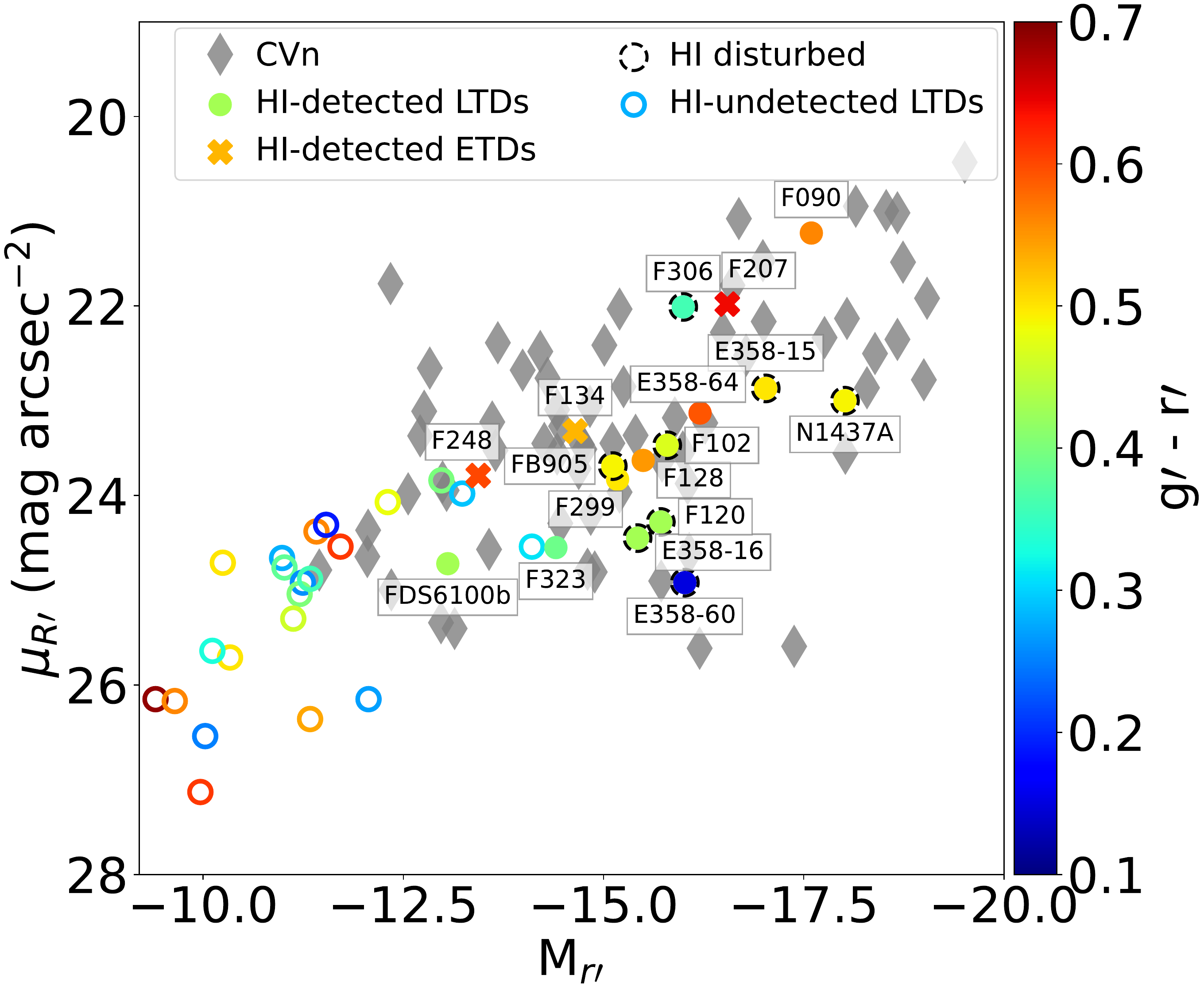}

\caption{Luminosity-projected surface brightness relation. The markers and colours are the same as Fig. \ref{fig:lum_HI}. The surface brightness for all Fornax LTDs and \HI-detected ETDs fall within the scatter of the CVn dwarfs. The five \HI-undetected LTDs between $M_{r'} = -12$ and $-14.2$ show typical surface brightness values for their luminosity.}

\label{fig:lum_surfbright}
\end{figure}

\begin{figure*}

    \includegraphics[width = \textwidth]{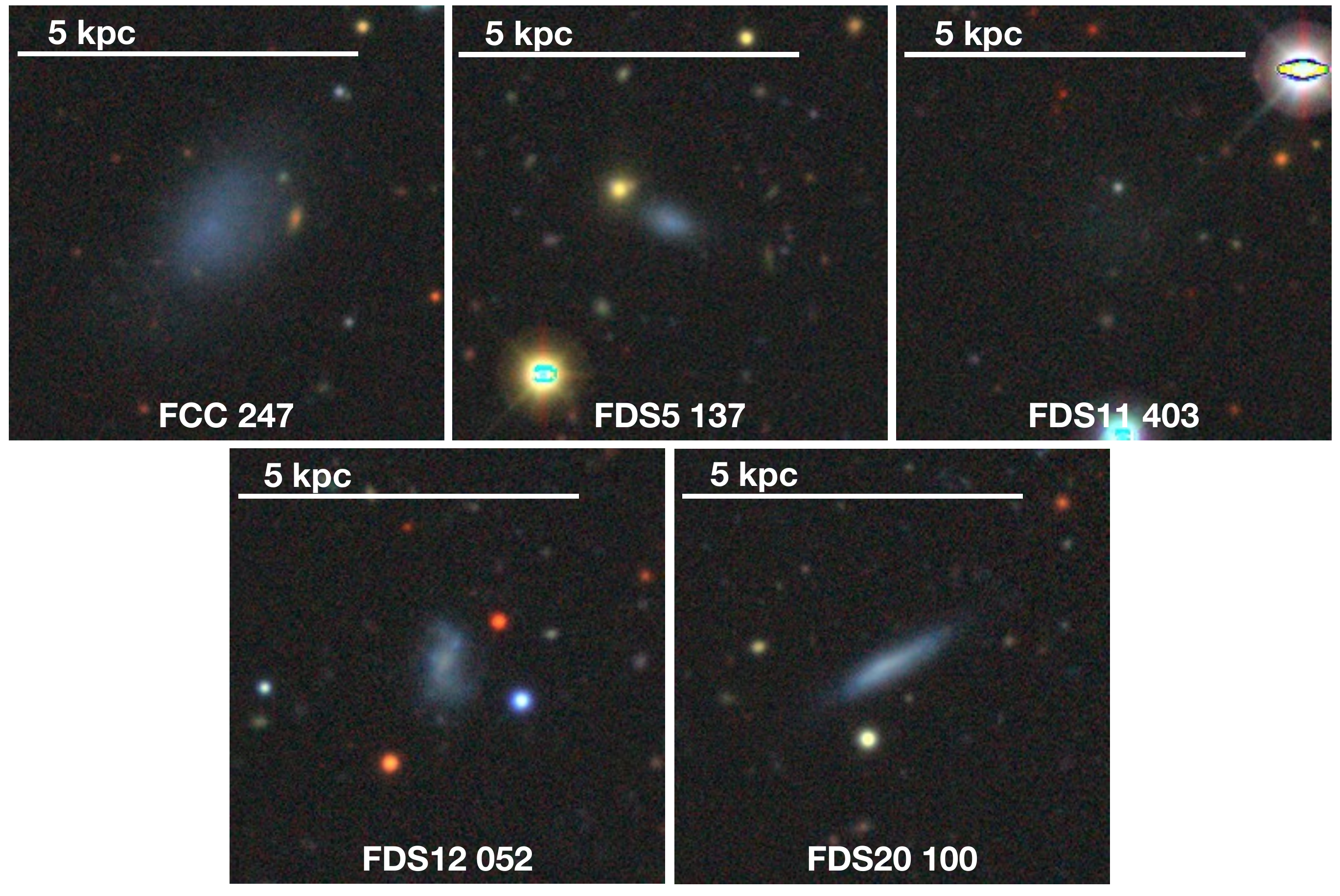}

\caption{DECaLS three-colour image of the \HI-undetected LTDs. The optical three-colour image for the five blue, LTDs between $M_{r'} = -12$ and $-14.2$ that we would expect to detect in \HI. Their colour, luminosity, morphology and surface brightness is the same as \HI-detected LTDs, suggesting that these five LTDs have very recently lost their \HI.}

\label{fig:blue_noHI}
\end{figure*}

We measure the projected surface brightness within the effective radius ($\mu_\mathrm{e}$) of the CVn dwarfs and compare it to that of the Fornax LTDs \citep{Su2021}. As above, we also include the \HI-detected ETDs in this comparison, even though we expect them to be a different, peculiar type of object. In Fig. \ref{fig:lum_surfbright}, we can see that the LTDs and \HI-detected ETDs lie within the scatter of the $M_{r'}$--$\mu_\mathrm{e}$ relation of the CVn galaxies. The five \HI-undetected, blue LTDs between $M_{r'} = -12$ and $-14.2$ show typical surface brightness values for their luminosity. They have low stellar masses that range (0.3 -- 1.6) $\times$ 10$^{7}$ \Msol, making it likely that their \HI has been very recently stripped such that neither the stellar structure nor blue stars have had a chance to fade yet. We show the DECaLS 3-colour optical images for these LTDs in Fig. \ref{fig:blue_noHI}. 

\subsection{Cluster substructure and dwarf local environment}
\begin{figure*}

    \includegraphics[width = \textwidth]{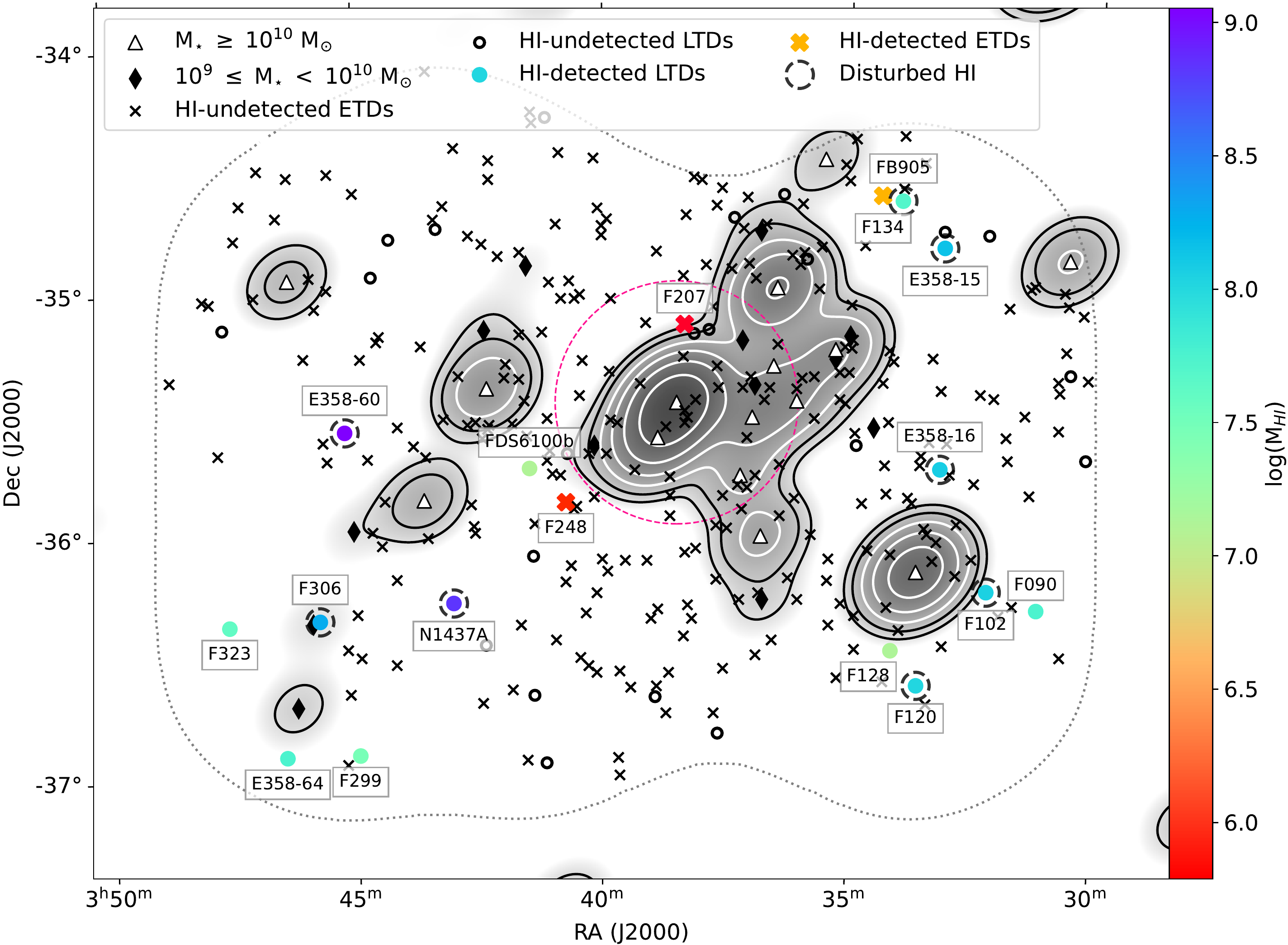}

\caption{Fornax cluster projected density field. The background image shows the projected density field on a log scale, produced by Gaussian smoothing the luminosity ratio relative to the BCG (NGC\,1399), at the position of each galaxy. The (black and white) contours emphasise the density field, where the lowest shows the influence of \Mstar $\simeq$ 5 $\times$ 10$^{9}$ \Msol and increases by 2$^{n}$. The grey dotted line shows the border of our observed area (same as Fig. \ref{fig:dwarf_locations}) the pink dashed circle is 0.25\Rvir. Black crosses and open black circles are the \HI-undetected ETDs and LTDs. The filled circles and thick crosses are the \HI-detected ETDs and LTDs, which have been labelled and coloured by their \MHI. LTDs with disturbed \HI morphologies are enclosed with dashed black circles. Black diamonds show the intermediate galaxies and the white triangles are the massive galaxies. The projected substructure is dominated by the brightest (hence most massive) galaxies, and the \HI-detected dwarfs avoid the most massive potentials (i.e. cluster centre and massive galaxies) in the cluster.}

\label{fig:num_dens}
\end{figure*}

The number density (i.e. local environment) of galaxies is crucial in shaping their evolution. The mass of the nearby galaxies (i.e. central vs satellite) also plays a vital role in \HI removal and star formation quenching \citep[e.g.][]{Spekkens2014, Poulain2022, Zhu2023}. With this in mind, we create a projected density field to reveal the substructure of the cluster. At the central coordinates of each galaxy in our field, we assign its relative luminosity, normalised to the BCG (NGC\,1399), and zero everywhere else. The image is then smoothed by a Gaussian kernel to produce the projected density field shown in Fig. \ref{fig:num_dens}. The normalised luminosity factors in brighter (i.e. massive) galaxies having a stronger influence on nearby galaxies than the standard projected density (e.g. $\Sigma_{5}$). Therefore, this projected density field is a proxy for the cumulative, stellar potential throughout the cluster. In Fig. \ref{fig:num_dens}, the massive galaxies (white triangles) are clearly contributing the most to the density field. Easily identifiable features include the BCG (at the centre of the pink circle showing 0.25\Rvir), the concentration of bright early-type galaxies west of it, and the bright spiral NGC\,1365 towards the south-west.

Figure \ref{fig:num_dens} shows that \HI-detected dwarfs avoid the most massive potentials, whether that be the cumulative substructure in the cluster centre (and just west of it), or individual massive galaxies. This is consistent with the well-known morphology-density relation \citep{Dressler1980} and with the Next Generation Fornax Survey \citep{Munoz2015}, who show a higher number of nucleated (i.e. more evolved) dwarfs in the cluster centre \citep{Ordenes-Briceno2018a, Ordenes-Briceno2018b}. The dwarfs with \HI are typically situated at large clustocentric radii (i.e. the majority are beyond 0.5\Rvir), which holds true for 13 out of the 14 \HI-detected LTDs and is what we would expect for LTDs on their first infall. Given that the LTDs with disturbed \HI are also far from the cluster core, their \HI is being actively shaped well before they reach the cluster centre. This is consistent with the star-forming dwarfs being concentrated in the outer regions of the cluster, shown by \citet{Drinkwater2001}. There is a single \HI-detected ETD (FCC\,207) near the cluster centre, $<$ 0.25\Rvir in projection. This is a peculiar case and we discuss it in the next section.

\begin{figure}

    \includegraphics[width = \columnwidth]{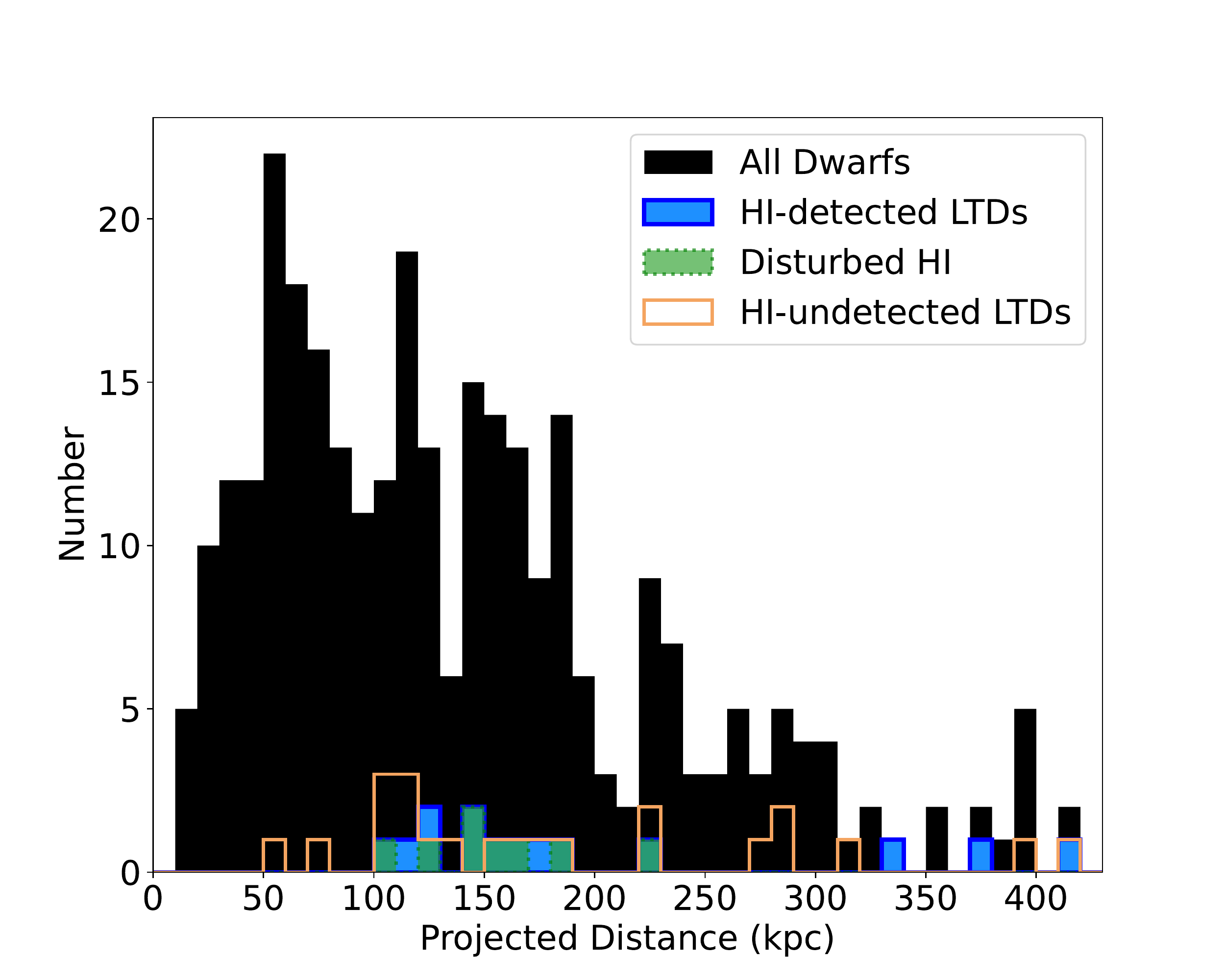}

\caption{Projected distance to nearest massive galaxy. All dwarfs in our observed area are shown as the filled-black histogram. \HI-detected LTDs are shown in the blue-filled histogram, LTDs with disturbed \HI are shown by the dashed green-filled histogram and \HI-undetected LTDs are the open orange histogram. There are no \HI-detected LTDs within a projected distance of 100\,kpc to their nearest massive galaxy, implying that massive galaxies play an active role in the removal of \HI from LTDs.}

\label{fig:proj_dist}
\end{figure}

We quantify the projected distance to the closest massive galaxy for all the LTDs in Fig. \ref{fig:proj_dist}. The \HI-detected LTDs are situated at least 100\,kpc from their nearest massive galaxy in projection. There are two \HI-undetected LTDs within 100\,kpc of their nearest massive galaxy. The LTDs with disturbed \HI tend to reside closer (e.g. between 100 and 190\,kpc) to their nearest massive galaxy when compared to the LTDs with undisturbed \HI. While this is not definite evidence, and projection effects introduce a lot of uncertainty, it does support the notion that LTDs can be stripped of their \HI by central galaxies and maintain their structure for a short time. Previous studies have found gas poor and quenched dwarf satellites in low-density environments (such as the local group) and have suggested that centrals have a universal and effective satellite quenching mechanism, such as ram-pressure stripping by the central halo \citep[e.g.][]{Poulain2022, Zhu2023}. 

\section{The mechanisms and timescale of \HI removal}
\label{sec:discussion}
There are multiple pieces of evidence that suggest the dwarf galaxies rapidly lose their \HI in the Fornax cluster. Firstly, the most abundant galaxies in the Fornax cluster are red, ETDs \citep{Venhola2018, Venhola2019}. However, the cluster is still assembling, which is evident by the substructure in Fig. \ref{fig:num_dens} and the un-settled giant early types, located in a region where many galaxies host asymmetric halos and low surface brightness features \citep{Iodice2016, Iodice2019, Spavone2020, Spavone2022}. Furthermore, there are different populations of recent infallers; massive late-types that are gas-rich and have disturbed stellar bodies, moderate atomic and molecular gas ratios and star formation rate deficiencies \citep{Raj2019, Zabel2019, Loni2021, Morokuma-Matsui2022}, and a range of star-forming dwarfs, located at the cluster outskirts \citep{Drinkwater2001}. The dwarfs with redshifts have a high velocity dispersion compared to both the cluster and the giants, implying that dwarfs are also currently infalling into the cluster with $\sim$70\% on radial orbits \citep{Drinkwater2001, deRijcke2010, Maddox2019}.

As described in Section \ref{sec:results}, we detect \HI in 14 out of the 36 LTDs, including all LTDs brighter than $M_{r'} = -14.2$. Interestingly, all \HI-detected LTDs contain the expected amount of \HI as LTDs at the same luminosity in the field (e.g. Fig. \ref{fig:lum_HI}). Because the \HI-detected LTDs have a normal \MHI and they are located at large clustocentric distances (Fig. \ref{fig:num_dens}), they must have recently joined the cluster and are on their first infall. Furthermore, eight of the LTDs show disturbed or asymmetric \HI morphologies. Considering that small perturbations, even from secular processes such as stellar feedback are able to induce asymmetries in the stellar and \HI disk in field LTDs \citep[e.g.][]{Begum2008, Hunter2012, Zhang2012, Poulain2022}, it is unsurprising that the majority of \HI-detected LTDs in Fornax also display stellar and \HI asymmetries. The Fornax LTDs are subjected to the same secular processes as those in the field, as well as interacting with the ram-pressure wind of the ICM and massive halos, tidal interactions with nearby galaxies and the potential of the cluster. The Fornax \HI-detected LTDs are analogous to field LTDs; although, a clear difference is the unambiguous evidence of the ICM ram-pressure shaping the outer \HI disk of some LTDs. 

\citet{Serra2023} showed that NGC\,1437A and FCC\,306 have starless, \HI tails pointing away from the cluster centre, with velocity gradients aligned with the direction of the ram-pressure wind. Using higher resolution \HI images (Fig. \ref{fig:overlays}), we show that the \HI on the side facing the cluster centre is being compressed into the stellar disk, an essential ingredient required to disentangle ram-pressure and tidal tails. We further show another three disturbed \HI-detected LTDs (ESO\,358-G016, FCC\,102 and FCCB\,905) that may also show evidence of ram-pressure shaping as they have potential \HI compressions and tails pointing away from the cluster centre. \citet{Serra2023} showed that a combination of ram pressure and tidal forces are required to operate together in the Fornax cluster. First, the \HI is pulled out of the stellar body by tidal forces. Then, this weakly bound \HI is further displaced and shaped by ram pressure as the galaxies travel through the ICM \citep{Serra2023}. This is consistent with the low ICM density and high number density of the Fornax cluster along with the asymmetric stellar bodies of the \HI-disturbed LTDs. \citet{Raj2019} noted the irregular stellar body of NGC\,1437A and suggested that the disturbed stellar emission in FCC\,306, ESO\,358-G015, ESO\,358-G016 and ESO\,358-G060 are due to interactions with the cluster potential. This is further supported by the simulations of \citet{Mastropietro2021} who showed that the tidal field of a Fornax-like cluster is able to cause damage to the stellar disks of low-mass galaxies and that perpendicular gaseous and stellar tails are explainable if they are subjected to different environmental effects. Due to the high galaxy number density in Fornax and influence of massive galaxies (Figs \ref{fig:num_dens} and \ref{fig:proj_dist}), we cannot rule out tidal interactions between galaxies as a mechanism to distort the stellar bodies. For example, FCC\,306 is located next to NGC\,1437B in projection, an intermediate mass galaxy. Even though both have disturbed stellar disks, they are separated by 700 \kms in velocity and there is no \HI connecting them down to a column density of 10$^{18}$ cm$^{-2}$. Another example of tidal interactions between galaxies is that three (ESO\,358-G016, FCC\,102 and FCC\,120) of the five \HI-detected LTDs satellites of NGC\,1365 have disturbed \HI disks and stellar bodies. As previously suggested by \citet{Loni2021}, these LTDs may be subjected to both pre-processing and early interactions with the cluster. 

Regarding the \HI-undetected LTDs, 17 out of 22 are fainter than $M_{r'} = -12$ and they lie within the 3$\sigma$ scatter of the CVn luminosity-\HI relation (Fig. \ref{fig:lum_HI}). They also have reasonable mean surface brightness values that follow the trend in Fig. \ref{fig:lum_surfbright}. These LTDs are so faint that even if they do contain \HI along the extrapolated CVn relation, it is below our detection limit. We would expect to detect \HI in the five remaining LTDs (between $M_{r'} = -12$ and $-14.2$) given their morphology, colour, luminosity and surface brightness. However, they all lie $>$3.5$\sigma$ below the CVn fit (Fig. \ref{fig:lum_HI}) and would need between $\sim$18 and 58 times more \HI to lie on the median of the relation. Fig. \ref{fig:blue_noHI} shows their 3-colour image of these LTDs and they visually appear similar (e.g. blue, with structure and some asymmetric stellar emission) to some of the \HI-detected LTDs. FDS11\,403 has a significantly lower mean surface brightness ($\mu_\mathrm{e}$ = 26.15) than the \HI-undetected LTDs in this luminosity range, and its diffuse nature is a plausible explanation for why no \HI is detected. The other four \HI-undetected LTDs may have very recently had their \HI stripped and not enough time has passed for them to lose their structure and have the blue stars fade to red. Their stellar masses range between 0.3 and 1.6 $\times$ 10$^{7}$ \Msol. A single shock to galaxies in this mass range can directly strip the \HI and initiate the quenching process. Therefore, low-mass LTDs with \HI may not have a chance to display a disturbed \HI morphology given how rapidly it can be removed. Similarly, the \HI-disturbed LTDs in Fornax are the more massive LTDs. Their irregular \HI morphology will be visible for longer timescales as it will take a longer (although still rapid) time or multiple events to strip all of the \HI. 

FCC\,090 is an interesting case. While it is classified as a late type \citep{Venhola2018}, there are significant differences between it and the other \HI-detected LTDs. It has the $\sim$ same brightness and stellar mass as NGC\,1437A, although FCC\,090 has a lower amount of \HI by more than a factor of 10 (Table \ref{tab:gen_prop}). FCC\,090 is one of the least blue \HI-detected LTDs, occupying the $\sim$ same colour region as the ETD FCC\,134 (Figs. \ref{fig:overlays} and \ref{fig:CMR}). FCC\,090 has a very high surface brightness (Fig. \ref{fig:lum_surfbright}) and is not completely smooth, with a potential bar in the core. However, its stellar body is significantly smoother than the other \HI-detected LTDs. It resides on the edge of the scatter in the CVn luminosity-\HI fit (Fig. \ref{fig:lum_HI}) and the velocity field (Fig \ref{fig:overlays}) does not show any rotation, which is akin to the ETDs with \HI. A tail of CO extending beyond the stellar body and pointing away from the cluster centre has been detected in FCC\,090 \citep{Zabel2019}. FCC\,090 has a low molecular gas fraction and ram-pressure was suggested to have created the unusual CO tail \citep{Zabel2019}, however, our \HI image (Fig. \ref{fig:overlays}) does not show a ram-pressure tail. This does not exclude ram-pressure previously stripping some \HI gas and the mixed LTD and ETD characteristics suggest that FCC\,090 is a transition type dwarf \citep[TTD;][]{Koleva2013}. Even though the \HI and CO have not been fully stripped, there is a small amount of \HI in the core above a column density of 10$^{21}$ cm$^{-2}$. FCC\,090 has a blue core that coincides with the high-density \HI, although there is no dense \HI outside this region to collapse and form stars. This may explain why the stellar morphology is smoother than the other \HI-detected LTDs. As FCC\,090 is being stripped of gas while the blue stars and structure are fading, the morphological transformation may occur on a similar timescale to the stripping of cool gas. 

ESO\,358-G064 is similar to FCC\,299 as they both are edge-on LTDs with \HI that does not deviate much from symmetric emission. Both galaxies also display a regularly rotating disk in the velocity field (Fig. \ref{fig:overlays}). The clear difference is that FCC\,299 is blue (as expected) and ESO\,358-G064 is not. ESO\,358-G064 is as red as FCC\,090 and redder than the ETD FCC\,134 (Figs. \ref{fig:overlays} and \ref{fig:CMR}). A red colour can indicate the quenching of star-formation, which may be a consequence of pre-processing. However, no other observations support this as ESO\,358-G064 is beyond \Rvir with typical \MHI and $\mu_\mathrm{e}$ values for its luminosity (Figs. \ref{fig:lum_HI}, \ref{fig:lum_surfbright} and \ref{fig:num_dens}). The neighbouring galaxy is a confirmed background source (Fig. \ref{fig:overlays}), ruling out tidal interactions.  Rather, in the case of ESO\,358-G064 the red colour is likely a result of the presence of dust. Dust can obscure stellar emission and artificially redden the colour of a galaxy. Most dwarf galaxies are not dusty and the internal extinction is often insignificant \citep[see][and referernces therein]{Henkel2022}. FCC\,090 FCC\,306 and NGC\,1437A have dust masses between 0.3 and 32 $\times$ 10$^{5}$ \Msol \citep{Zabel2021}. The optical image shows that ESO\,358-G064 does indeed contain some dust. As it edge-on, the dust obscuration can have an even higher internal extinction than face-on LTDs such as NGC\,1437A. A small (e.g. $\lesssim$ 0.07) colour correction would mean that the true colour of ESO\,358-G064 is the $\sim$ same as other LTDs with a comparable \MHI. Internal extinction from dust reddening is a more plausible explanation for the red colour of ESO\,358-G064 than the star-formation being quenched.

The dense (e.g. \NHI$\geq$10$^{21}$ cm$^{-2}$) \HI is required to be stripped from a LTD so that star formation ceases. Only then can the blue stars and structure fade, thus transforming into an red, quiescent ETD. Therefore, a morphological transformation from a LTD to an ETD cannot be faster than the dense \HI stripping time. However, as suggested above, it can be on a similar timescale, particularly if there are external tidal forces disrupting the stellar body. A possible explanation for the three \HI-detected ETDs is that they are the end phase of quenched LTDs. If a LTD is on a tangential orbit, the \HI will be slowly consumed to form stars. There may only be a small amount of low column density \HI remaining that is not dense enough to form stars by the time they are ETDs. Another scenario is that these ETDs have recently acquired some \HI through a `rejuvenation' event. An example of this could be a minor merger between a quenched ETD with a LTD containing some \HI. The ETD will appear blue from a  merger-induced starburst and retain a small amount of \HI. This has already been observed in the Fornax\,A subgroup \citep{deRijcke2013, Kleiner2021} and is indeed a plausible scenario in the Fornax cluster. FCC\,134 is the bluest (Fig. \ref{fig:CMR} and Table \ref{tab:gen_prop}) \HI-detected ETD and has been observed in the SAMI-Fornax Dwarfs survey \citep{Scott2020}. The average age of the stellar population of FCC\,134 is 3\,Gyr \citep{Romero-Gomez2023}. This is young for a ETD and is only possible if there has been recent star formation. FCC\,248 is located at the end of the \HI tail of NGC\,1427A. It is plausible that it overlaps in projection and does not actually contain \HI. However, at high resolutions the \HI column density peaks exactly at the position of FCC\,248 and the radial velocities of FCC\,248 and the surrounding \HI emission from the NGC\,1427A tail are the same. Rather than being a projection effect, we believe that FCC\,248 is a quenched, red ETD that accreted some \HI from the tail of NGC\,1427A, but is not dense enough to form new stars. FCC\,207 is located near the cluster core, on the edge of the densest region of the cluster (Figs. \ref{fig:dwarf_locations} and \ref{fig:num_dens}). It is difficult to explain how this red ETD has any \HI and off-centre molecular gas with disturbed kinematics \citep{Zabel2019}. The average age of the stellar population is 6\,Gyr \citep{Romero-Gomez2023}, which does not support any recent star formation. There may be projection effects regarding its true location, although the velocity of FCC\,207 is the same as the systemic cluster velocity. Perhaps FCC\,207 favours the tangential orbit slow quenching scenario \citep[similarly suggested by][in the Virgo cluster]{Conselice2003b} or it is not actually that near to the cluster core.

\section{\HI removal toy model}
\label{sec:toymodel}
The cluster crossing time is $\sim$2\,Gyr \citep{Loni2021} and active gas removal that quench dwarf galaxies on quicker timescales than the crossing time has already been proposed \citep{deRijcke2010, Drinkwater2001, Loni2021}. Similarly, previous works have suggested that dwarf galaxies will lose all of their \HI soon after the crossing of the virial radius or at first pericentric passage in other (e.g. Virgo) clusters \citep{Boselli2008, Boselli2014b, Hester2010, Cortese2021}. This is very difficult to show observationally, as we do not have star formation histories (SFHs) for our sample of dwarfs, which can be used to constrain the gas removal and quenching timescales. Thanks to our very sensitive and deep MeerKAT observations, we are confident about which galaxies have \HI down to 5 $\times$ 10$^{5}$ \Msol. Therefore, we construct a simple toy model, that models the \HI loss as dwarf galaxies fall into the Fornax cluster, to reproduce what we observe in Fig. \ref{fig:lum_HI}.

In this toy model, we control only 3 parameters; (i) The accretion rate ($N_\mathrm{infall}$) of dwarf galaxies falling onto the cluster i.e. number of accreted dwarfs per Gyr, (ii) $t_{1/2}$, which is the halving time of \HI, assuming an exponential \MHI loss using $M_\mathrm{HI}(t) = M_\mathrm{HI, 0} e^{\lambda t/t_{1/2}}$ and $\lambda$ = $ln(0.5)$. Lastly (iii) the accretion clustering i.e. how grouped together the dwarfs being deposited onto the cluster are -- no clustering would be individual galaxies deposited one-at-a-time and some clustering is when multiple dwarfs are deposited simultaneously. 

Galaxies in the field that consume \HI to form stars will have a $\sim$ constant \MHI over cosmic time, provided they have a supply of gas from the cosmic web. Strong environmental encounters are responsible for removing significant amounts of \HI. Hydrodynamical simulations \citep[e.g.][]{Morasco2016, Rohr2023} have shown that ram pressure, tidal interactions and the combined effect of both, rapidly removes a significant amount of \HI. After this initial phase of \HI loss, the remaining \HI will be removed at a lower rate as there is less \HI to remove and it resides deeper in the potential well. We approximate this process to first order in our toy model by using an exponential decrease of \MHI over time.

To reproduce Fig. \ref{fig:lum_HI}, we impose the same observational constraints we observe in Fornax; we do not detect \HI in dwarfs fainter than $M_{r'} = -13$ nor do we detect any \HI below \MHI = 5 $\times$ 10$^{5}$ \Msol (Fig. \ref{fig:lum_HI}). We are qualitatively looking to reproduce two fundamental things; the number of LTDs within the 3$\sigma$ scatter of the CVn fit, which should be of the order of $\sim$ 10, and numerous objects below the detection limit, though in practice we accept having a few objects just above it to account for the low number statistics and error bars. If either of these criteria are not satisfied, then there will be too few or too many LTDs on the CVn fit or numerous detections between \MHI = 5 $\times$ 10$^{5}$ \Msol and the $M_{r'}$--\MHI relation, which are not observed in Fornax. We presume there is nothing special about observing the Fornax cluster at this time, and given that we are modelling the \MHI loss, we should be able to reproduce Fig. \ref{fig:lum_HI} in multiple snapshots. 

\begin{figure*}

    \includegraphics[width = \textwidth]{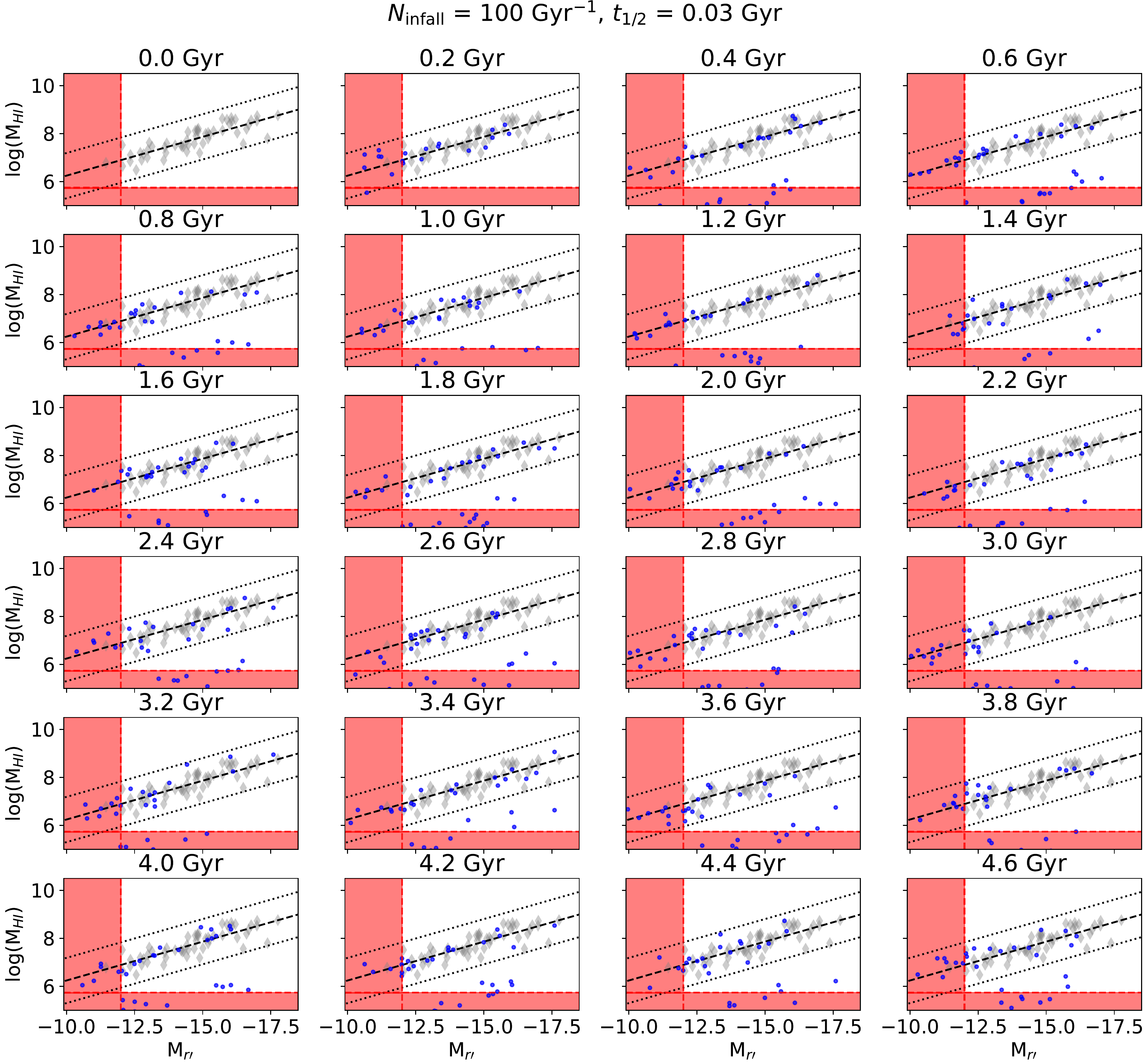}

\caption{Toy model of \HI removal in the cluster. The best results to reproduce Fig. \ref{fig:lum_HI}, showing what we would observe every 0.2\,Gyr. The grey diamonds are the CVn galaxies, the dashed black line shows the linear fit used to populate infalling galaxies to the CVn galaxies and the dotted lines show the 3$\sigma$ scatter. The blue filled circles are galaxies in the cluster and the red dashed lines are the observational constraints both in $M_{r'}$ and \MHI, such that we would not detect a galaxy in \HI in the red shaded regions. With N$_{infall}$ = 100\,Gyr$^{-1}$ and $t_{1/2}$ = 0.03\,Gyr we are able to qualitatively reproduce what we observe in Fig. \ref{fig:lum_HI}.} 

\label{fig:toy_model}
\end{figure*}

We model infalling dwarf galaxies onto the cluster as blue, field LTDs that populate the $M_{r'}$ - \MHI plane according to the CVn distribution. To achieve this, a random FDS dwarf magnitude chosen and the corresponding \MHI is computed from a random normal distribution centred on the $M_{r'}$ - \MHI CVn fit with 3$\sigma$ scatter (Fig. \ref{fig:lum_HI}). Then, galaxies in the cluster exponentially lose their \MHI (as above) and additional dwarfs are continually deposited onto the cluster as time evolves. We explored a combination of parameters that spanned two orders of magnitude both in $N_\mathrm{infall}$ and $t_{1/2}$. In Fig. \ref{fig:toy_model} we show the result with the best values to reproduce Fig. \ref{fig:lum_HI}. A $N_\mathrm{infall}$ = 100\,Gyr$^{-1}$ is required to reproduce $\sim$ 10 (i.e not too few or too many) dwarfs on the CVn fit and is a reasonable value as we know that $\sim$ 300 dwarfs are in the Fornax cluster within the observed MFS area at $z$ $\sim$ 0. The best value for the \MHI halving time is $t_{1/2}$ = 0.03\,Gyr, as the dwarfs lose \HI rapidly enough that numerous objects are below the detection limit with few objects just above it at any given snapshot. However, we are only able to simultaneously reproduce both these criteria by including some accretion clustering that  simultaneously deposits 20 dwarfs onto the cluster. This is actually a more realistic scenario and akin to how the cosmic web funnels galaxy groups onto clusters. The majority of snapshots qualitatively reproduce what we observe in Fornax, although snapshots at t = 0.4, 1.0, 1.8, 2.8, 4.0 and 4.2 Gyr are the most alike Fig. \ref{fig:lum_HI}.

In Table \ref{tab:HI_loss}, we show the time it takes for galaxies in the \MHI range of 10$^{6}$--10$^{9}$ to be undetectable by the MFS in our toy model. This is a simplistic way of modelling the \HI loss as it does not take into account how galaxies lose their \HI, only the rate that \HI is removed. Therefore, the timescales in Table \ref{tab:HI_loss} should not be overstated. Despite the simplicity, our timescales are consistent with low-mass dwarfs being directly stripped of their \HI in a single shock (e.g. what we expect happened to the \HI-undetected LTDs between $M_{r'} = -12$ and $-14.2$), more massive LTDs harbouring disturbed \HI morphologies due to longer times or multiple events being required to fully strip their \HI, and not detecting any LTDs in \HI below the $M_{r'}$ - \MHI relation. These timescales are significantly shorter than the 2\,Gyr cluster crossing time and consistent with active and efficient \HI removal from LTDs in the Fornax cluster. In the Virgo cluster it has also been estimated that LTDs with \MHI $\sim$ 4 $\times$ 10$^{7}$ \Msol will be stripped of \HI in $\sim$ 150\,Myr \citep{Boselli2008, Kenney2014}. While we constrain the \HI removal timescale for dwarf galaxies in the Fornax cluster, the morphological transformation from a LTD to an ETD in clusters and the field is not well constrained. As mentioned previously, it cannot be quicker than the \HI removal timescale, although it may be able to occur on a similar timescale, particularly with external tidal fields shaping the stellar body. FCC\,090 may suggest that a morphological transformation can occur in parallel with the \HI removal, although this is just one case and not direct evidence of both transformations occurring at the same rate.

\begin{table}
  \begin{center}
    \caption{Approximate time it takes for dwarfs with different \MHI to be undetectable in the MFS, as measured from our \HI removal toy model.}
    \label{tab:HI_loss}
    \begin{tabular}{c c }
        \hline
        \hline
         log(\MHI) & $\sim$$t_\mathrm{\HI\,loss}$ \\
         (\Msol) & (Myr) \\
        \hline        
        6 & 30 \\
        7 & 150 \\ 
        8 & 240 \\
        9 & 330 \\
        \hline
    \end{tabular}
  \end{center}

\end{table}

\section{Summary}
\label{sec:conclusion}

We present results on the \HI content of dwarf galaxies in the central $\sim$$2.5 \times 4$ deg$^2$ of the Fornax cluster located 20\,Mpc away as part of the MeerKAT Fornax Survey. In this work we show three different \HI images that correspond to a 3$\sigma$ column density sensitivity range between 2.7 and 50 $\times$ 10$^{18}$\,cm$^{-2}$ over 25\,\kms for spatial resolutions between 4 and 1\,kpc. Such impressive sensitivity allows us to detect a 3$\sigma$ point source with a line width of 50\,\kms down to a \MHI = 5 $\times$ 10$^{5}$ \Msol.  

We detected \HI in 17 out of the 304 dwarfs. Seven are new \HI detections and we discovered new \HI features in previously known detections. Our \HI-detected dwarfs consist of 14 out of the 36 LTDs and three of the 268 ETDs. Eight of the \HI-detected LTDs show irregular or asymmetric \HI emission in the form of \HI tails, significant warps, or non-rotating components. NGC\,1437A and FCC\,306 are clearly having their \HI shaped by ram pressure as they have one-sided starless \HI tails pointing away from the cluster centre and velocity gradients aligned with the ram-pressure wind direction \citep{Serra2023}. We show in the high-resolution images that their \HI closest to the cluster centre is being compressed into the stellar disk along with an additional three LTDs that show evidence of \HI compression and one-sided tails, albeit they are less obvious than NGC\,1437A and FCC\,306. \citet{Serra2023} suggest that tidal forces are required in addition to ram pressure to reproduce the observed properties of some \HI-disturbed cluster galaxies and we observe the same as the eight \HI-disturbed LTDs have asymmetric or lopsided stellar emission. 

To reveal the substructure of the cluster, we created a projected density field and find that all the \HI-detected dwarfs avoid the most massive potentials (i.e. cluster centre and massive galaxies). None of the \HI-detected LTDs are within a projected distance of 100\,kpc to their nearest massive galaxy, implying that massive galaxies play an active role in the removal of \HI from LTDs. 

The \HI-detected LTDs have most likely recently joined the cluster and are on their first infall as they are located at large clustocentric radii, that is to say beyond 0.5\Rvir, and have comparable \MHI and $\mu_\mathrm{e}$ at fixed luminosity when compared to blue, star-forming LTDs in the field. We detected \HI in all LTDs brighter than $M_{r'} = -14.2$, which corresponds to a \Mstar between $\sim$ 1.6 $\times$ 10$^{7}$ and 10$^{9}$ \Msol, a range where LTDs can briefly display a disturbed \HI morphology. At luminositites between $M_{r'} = -12$ and $-14.2$, \Mstar range $\sim$(0.3 -- 1.6) $\times$ 10$^{7}$ \Msol, we detected one LTD with a normal amount of \HI, which is again analogous to a field LTD, and five LTDs that may have very recently had their \HI stripped. Given their colour, luminosity, morphology, and surface brightness, we would expect to detect \HI, but do not down to \MHI= 5 $\times$ 10$^{5}$ \Msol. Direct stripping from a single shock is likely to occur in this mass range (and below) such that these LTDs transition between having a normal amount of \HI to no \HI and never show a disturbed \HI morphology. No LTDs were detected in \HI fainter than $M_{r'} = -12$. Assuming that the $M_{r'}$ - \MHI relation can be extrapolated down to these low luminosities, our upper limits are consistent with those galaxies with a normal \MHI.
 
We created a simple toy model that deposits field LTDs onto the Fornax cluster and removes their \HI at an exponential rate. In order to reproduce what we observed in the Fornax cluster -- that is to say enough LTDs on the $M_{r'}$--\MHI plane and no LTDs detected in \HI below the $M_{r'}$--\MHI relation -- we find that the \HI of a \MHI = 10$^{8}$ \Msol dwarf is stripped in $\sim$240\,Myr. As this is much shorter than the cluster crossing time,  active \HI removal must be occurring in the cluster. Our toy model is consistent with low-mass LTDs being directly stripped of their \HI (e.g. through a single shock) and more massive LTDs harbouring disturbed \HI morphologies due to longer times or multiple events being required to fully strip their \HI. 

FCC\,090 is likely a TTD as it shares joint properties with LTDs and ETDs. It is one of the least blue with the smoothest stellar body of the \HI-detected LTDs, does not show any rotation in the \HI velocity field, and resides at the edge of the scatter below the median of the $M_{r'}$ - \MHI relation. This is tentative evidence that, in a cluster, a morphological transformation can occur in parallel on a similar timescale to the stripping of \HI. It is difficult to discern the origin of the peculiar \HI-detected ETDs. For two of them, we believe it is more likely that they have recently acquired some \HI through a `rejuvenation' event. FCC\,134 is blue with a young average stellar population age for an ETD and may have acquired some \HI through a minor merger that induced a starburst. FCC\,248 may have accreted some \HI from the tail of NGC\,1427A, but is not dense enough to form new stars. FCC\,207 does not fit the scenario of recently acquiring \HI. It is located near the cluster core with a typical ETD stellar population age. The systemic velocity of FCC\,207 is the same as the systemic cluster velocity; although, there still may be projection effects. Otherwise, a slow quenching process from a tangential orbit could potentially produce an ETD with a small amount of \HI near the cluster core. 

In this work, we have demonstrated how active and efficient \HI removal is required to reproduce the multi-wavelength properties of the dwarf galaxies in the Fornax cluster. The \HI is rapidly stripped from infalling galaxies of the order of a few hundred million years, which is responsible for producing the hundreds of quiescent ETDs in the cluster. The \HI is removed via both tidal and hydrodynamical forces operating in the cluster. While the MFS is ongoing, this work has covered the central and most efficient region of \HI-removal in the cluster, out to $\sim$\Rvir. This is the first time dwarf galaxies of this mass have been detected and resolved beyond the local group and in a galaxy cluster.

\begin{acknowledgements}
    
The MeerKAT telescope is operated by the South African Radio Astronomy Observatory, which is a facility of the National Research Foundation, an agency of the Department of Science and Innovation. We are grateful to the full MeerKAT team at SARAO for their work on building and commissioning MeerKAT. This work made use of the Inter-University Institute for Data Intensive Astronomy (IDIA) visualisation lab (https://vislab.idia.ac.za). IDIA is a partnership of the University of Cape Town, the University of Pretoria and the University of Western Cape.

This work has received funding from the European Research Council (ERC) under the European Union’s Horizon 2020 research and innovation programme (grant agreement No. 882793 "MeerGas"). This project has received funding from the European Research Council (ERC) under the European Union’s Horizon 2020 research and innovation programme (grant agreement no. 679627; project name FORNAX). MAR acknowledges funding from the European Union’s Horizon 2020 research and innovation programme under the Marie Skłodowska-Curie grant agreement No 101066353 (project ELATE). AL has received support from the Science and Technologies Facilities Council (STFC) through their New Applicant grant ST/T000503/1. FL acknowledges financial support from the Italian Minister for Research and Education (MIUR), project FARE, project code R16PR59747, project name FORNAX-B. FL acknowledges financial support from the Italian Ministry of University and Research $-$ Project Proposal CIR01$\_$00010. PK is supported by the BMBF project 05A20PC4 for D-MeerKAT. 

The Legacy Surveys consist of three individual and complementary projects: the Dark Energy Camera Legacy Survey (DECaLS; Proposal ID \#2014B-0404; PIs: David Schlegel and Arjun Dey), the Beijing-Arizona Sky Survey (BASS; NOAO Prop. ID \#2015A-0801; PIs: Zhou Xu and Xiaohui Fan), and the Mayall z-band Legacy Survey (MzLS; Prop. ID \#2016A-0453; PI: Arjun Dey). DECaLS, BASS and MzLS together include data obtained, respectively, at the Blanco telescope, Cerro Tololo Inter-American Observatory, NSF’s NOIRLab; the Bok telescope, Steward Observatory, University of Arizona; and the Mayall telescope, Kitt Peak National Observatory, NOIRLab. The Legacy Surveys project is honored to be permitted to conduct astronomical research on Iolkam Du’ag (Kitt Peak), a mountain with particular significance to the Tohono O’odham Nation.

NOIRLab is operated by the Association of Universities for Research in Astronomy (AURA) under a cooperative agreement with the National Science Foundation.

This project used data obtained with the Dark Energy Camera (DECam), which was constructed by the Dark Energy Survey (DES) collaboration. Funding for the DES Projects has been provided by the U.S. Department of Energy, the U.S. National Science Foundation, the Ministry of Science and Education of Spain, the Science and Technology Facilities Council of the United Kingdom, the Higher Education Funding Council for England, the National Center for Supercomputing Applications at the University of Illinois at Urbana-Champaign, the Kavli Institute of Cosmological Physics at the University of Chicago, Center for Cosmology and Astro-Particle Physics at the Ohio State University, the Mitchell Institute for Fundamental Physics and Astronomy at Texas A\&M University, Financiadora de Estudos e Projetos, Fundacao Carlos Chagas Filho de Amparo, Financiadora de Estudos e Projetos, Fundacao Carlos Chagas Filho de Amparo a Pesquisa do Estado do Rio de Janeiro, Conselho Nacional de Desenvolvimento Cientifico e Tecnologico and the Ministerio da Ciencia, Tecnologia e Inovacao, the Deutsche Forschungsgemeinschaft and the Collaborating Institutions in the Dark Energy Survey. The Collaborating Institutions are Argonne National Laboratory, the University of California at Santa Cruz, the University of Cambridge, Centro de Investigaciones Energeticas, Medioambientales y Tecnologicas-Madrid, the University of Chicago, University College London, the DES-Brazil Consortium, the University of Edinburgh, the Eidgenossische Technische Hochschule (ETH) Zurich, Fermi National Accelerator Laboratory, the University of Illinois at Urbana-Champaign, the Institut de Ciencies de l’Espai (IEEC/CSIC), the Institut de Fisica d’Altes Energies, Lawrence Berkeley National Laboratory, the Ludwig Maximilians Universitat Munchen and the associated Excellence Cluster Universe, the University of Michigan, NSF’s NOIRLab, the University of Nottingham, the Ohio State University, the University of Pennsylvania, the University of Portsmouth, SLAC National Accelerator Laboratory, Stanford University, the University of Sussex, and Texas A\&M University.

BASS is a key project of the Telescope Access Program (TAP), which has been funded by the National Astronomical Observatories of China, the Chinese Academy of Sciences (the Strategic Priority Research Program “The Emergence of Cosmological Structures” Grant \# XDB09000000), and the Special Fund for Astronomy from the Ministry of Finance. The BASS is also supported by the External Cooperation Program of Chinese Academy of Sciences (Grant \# 114A11KYSB20160057), and Chinese National Natural Science Foundation (Grant \# 11433005).

The Legacy Survey team makes use of data products from the Near-Earth Object Wide-field Infrared Survey Explorer (NEOWISE), which is a project of the Jet Propulsion Laboratory/California Institute of Technology. NEOWISE is funded by the National Aeronautics and Space Administration.

The Legacy Surveys imaging of the DESI footprint is supported by the Director, Office of Science, Office of High Energy Physics of the U.S. Department of Energy under Contract No. DE-AC02-05CH1123, by the National Energy Research Scientific Computing Center, a DOE Office of Science User Facility under the same contract; and by the U.S. National Science Foundation, Division of Astronomical Sciences under Contract No. AST-0950945 to NOAO.

\end{acknowledgements}

\bibliographystyle{aa} 
\bibliography{References} 

\begin{appendix}
\section{Full Fig. 3}
\label{sec:appen}

\begin{figure*}

    \includegraphics[width = \textwidth]{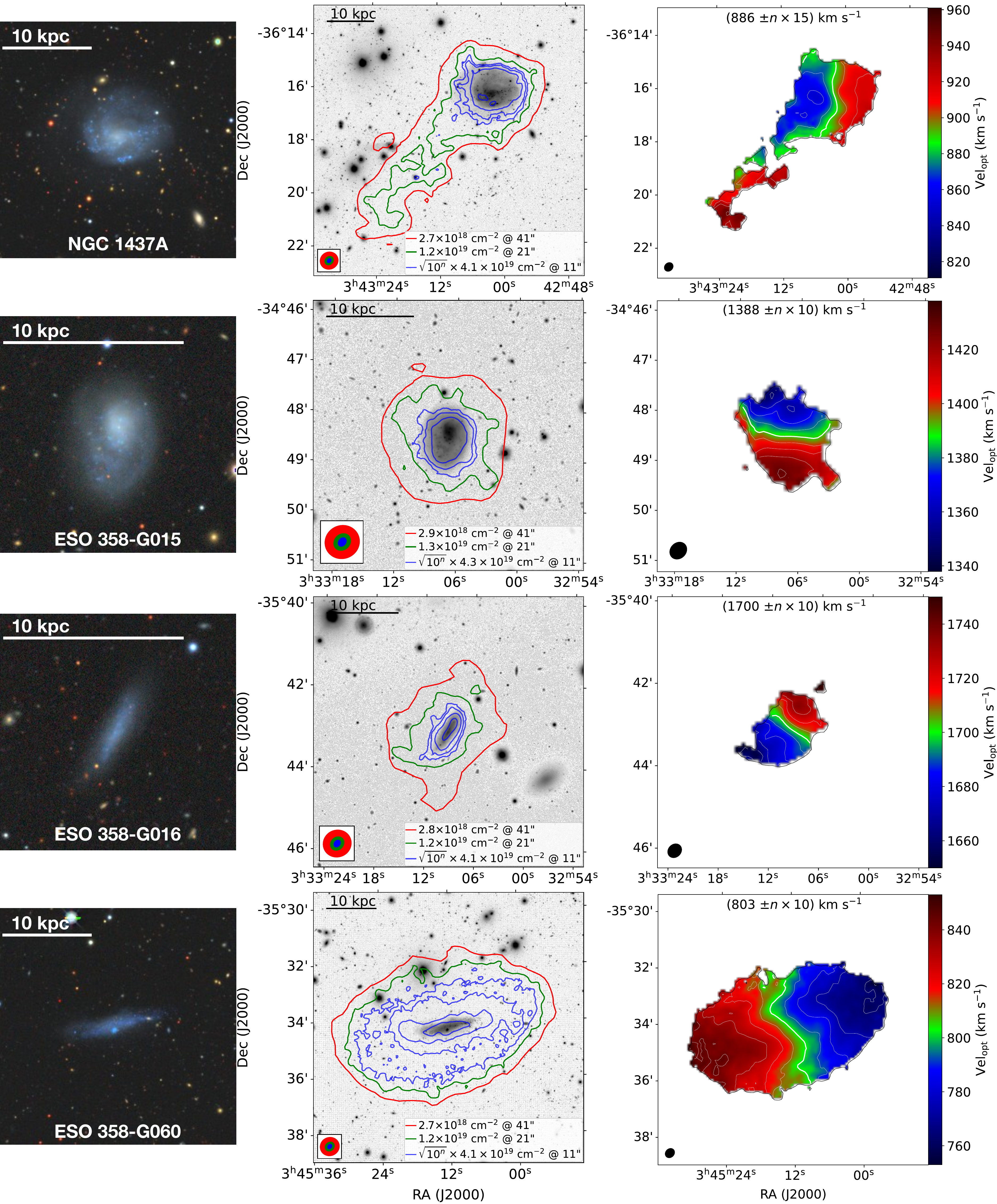}

\caption{Full Fig. \ref{fig:overlays_prev}, but displaying all 17 \HI detections.}

\label{fig:overlays}
\end{figure*}

\begin{figure*}

    \includegraphics[width = \textwidth]{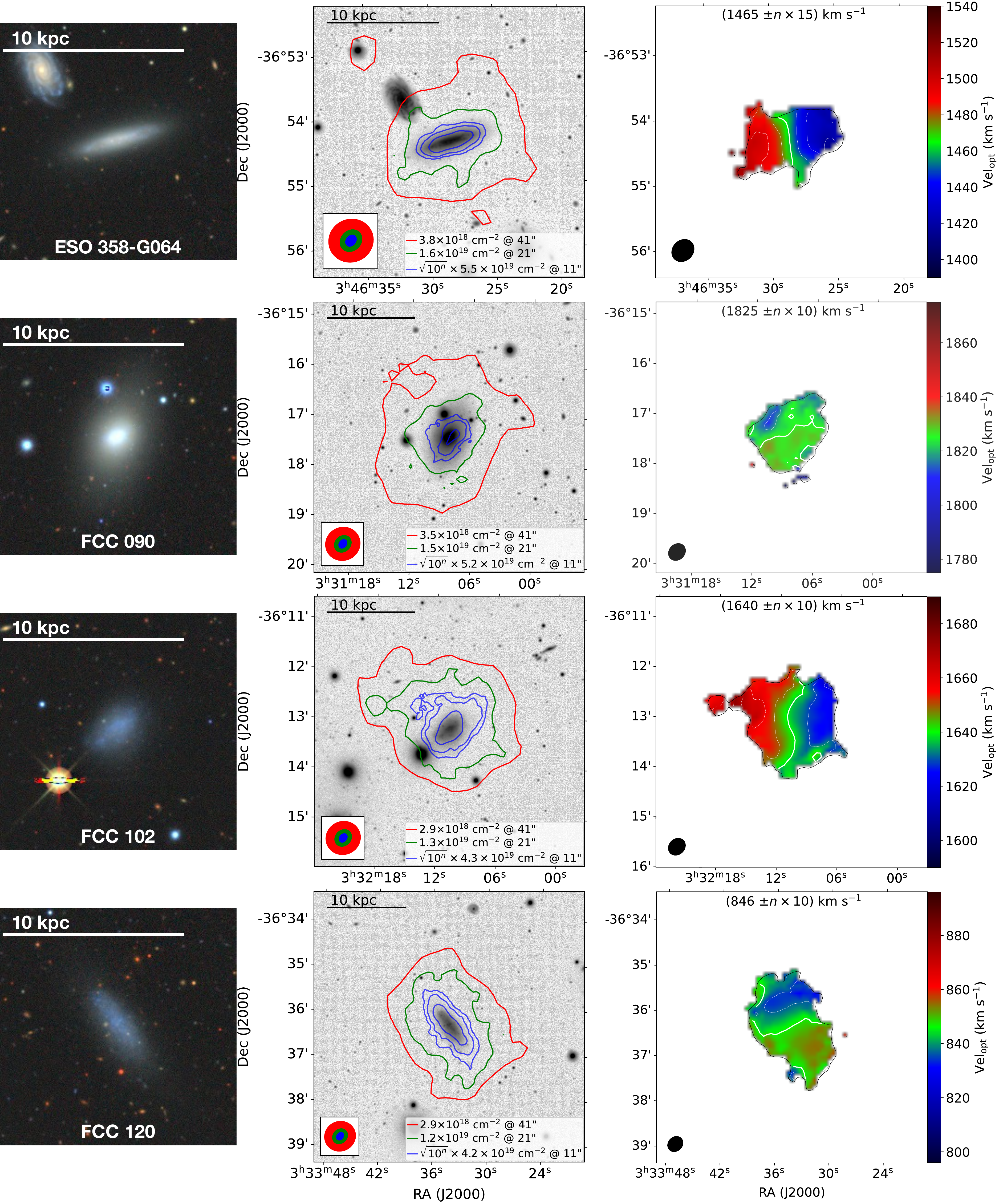}
    
  \addtocounter{figure}{-1}
   \caption{continued.}
\end{figure*}

\begin{figure*}

    \includegraphics[width = \textwidth]{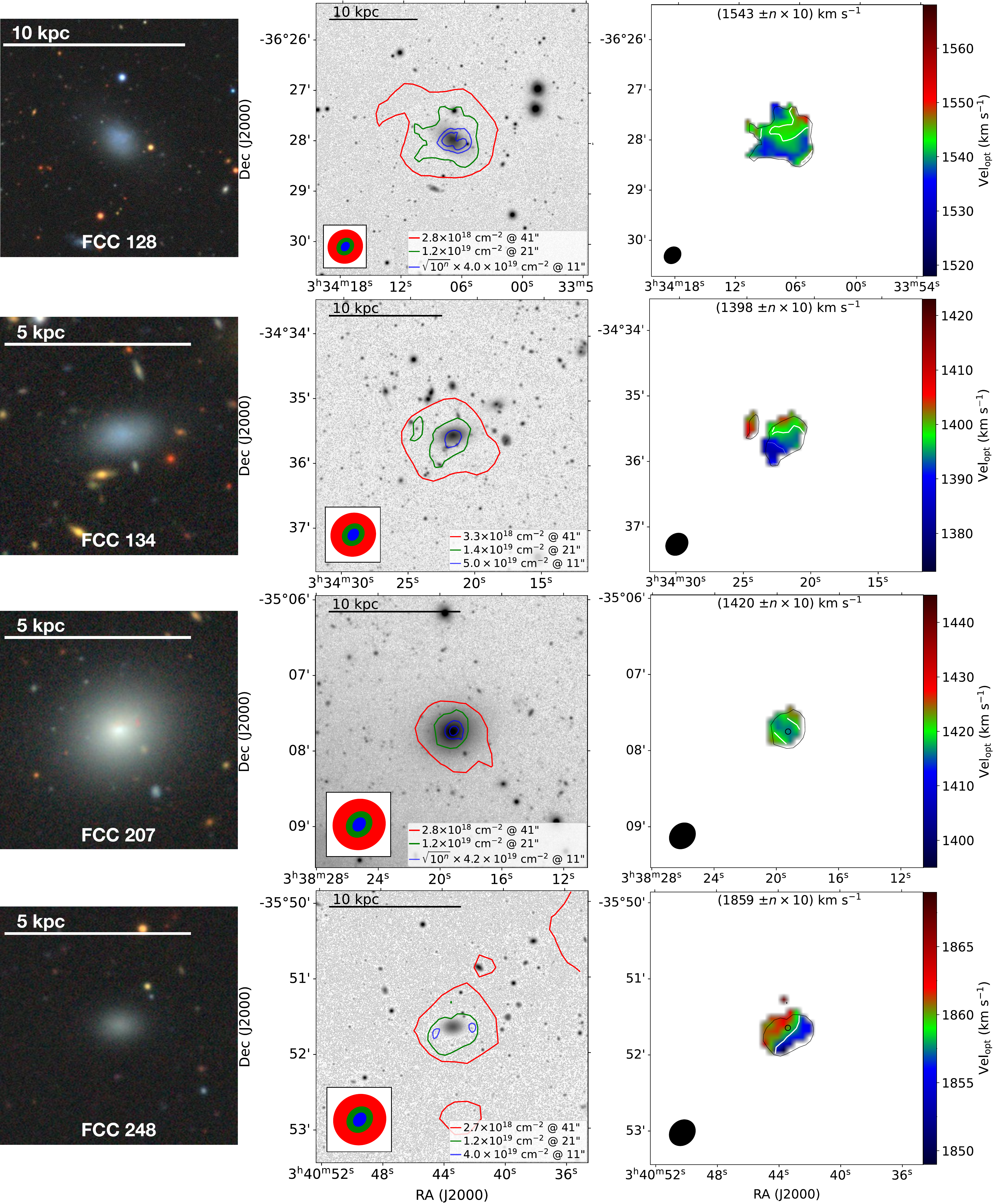}
    
  \addtocounter{figure}{-1}
   \caption{continued.}
\end{figure*}

\begin{figure*}

    \includegraphics[width = \textwidth]{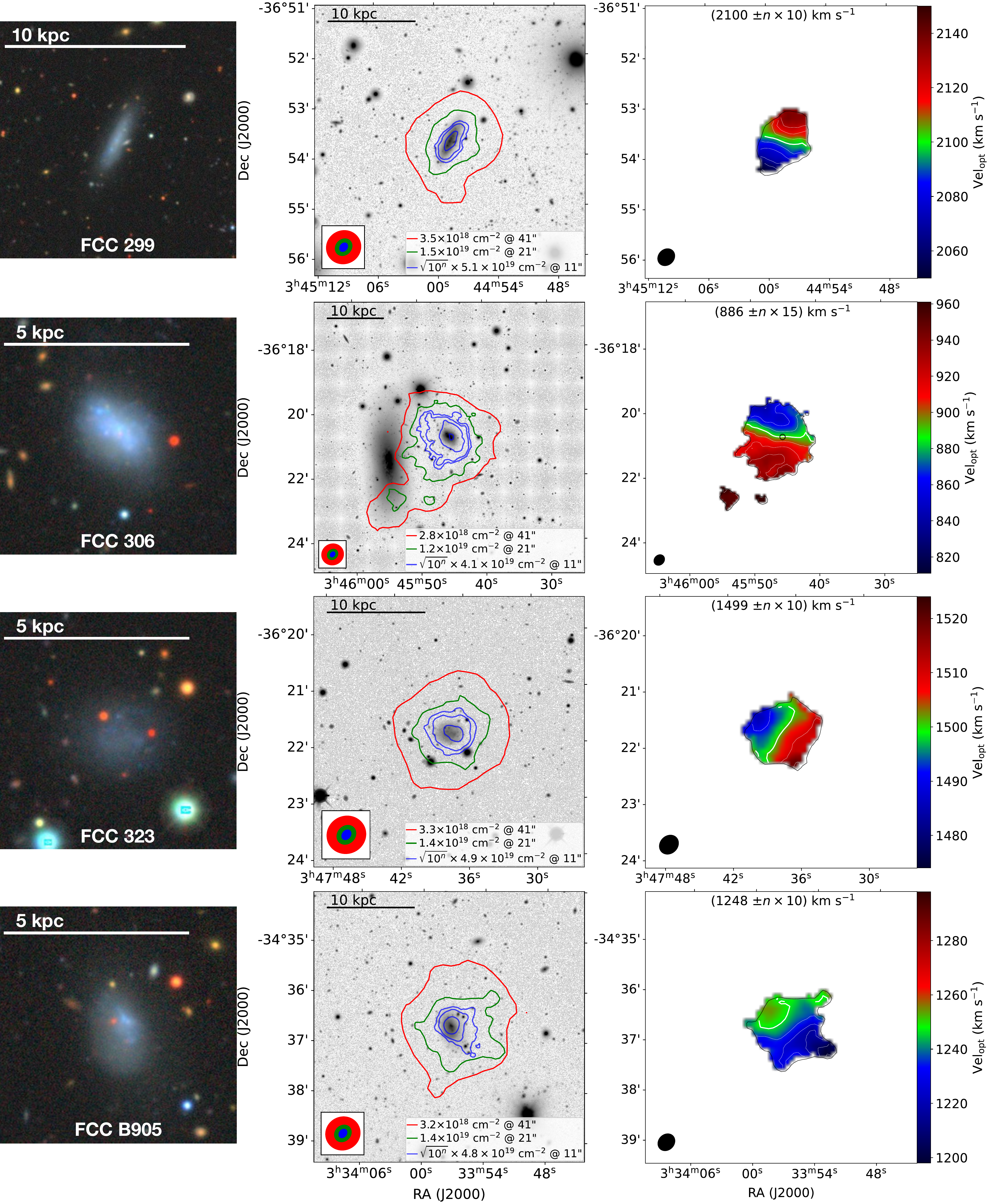}
    
  \addtocounter{figure}{-1}
   \caption{continued.}
\end{figure*}

\begin{figure*}

    \includegraphics[width = \textwidth]{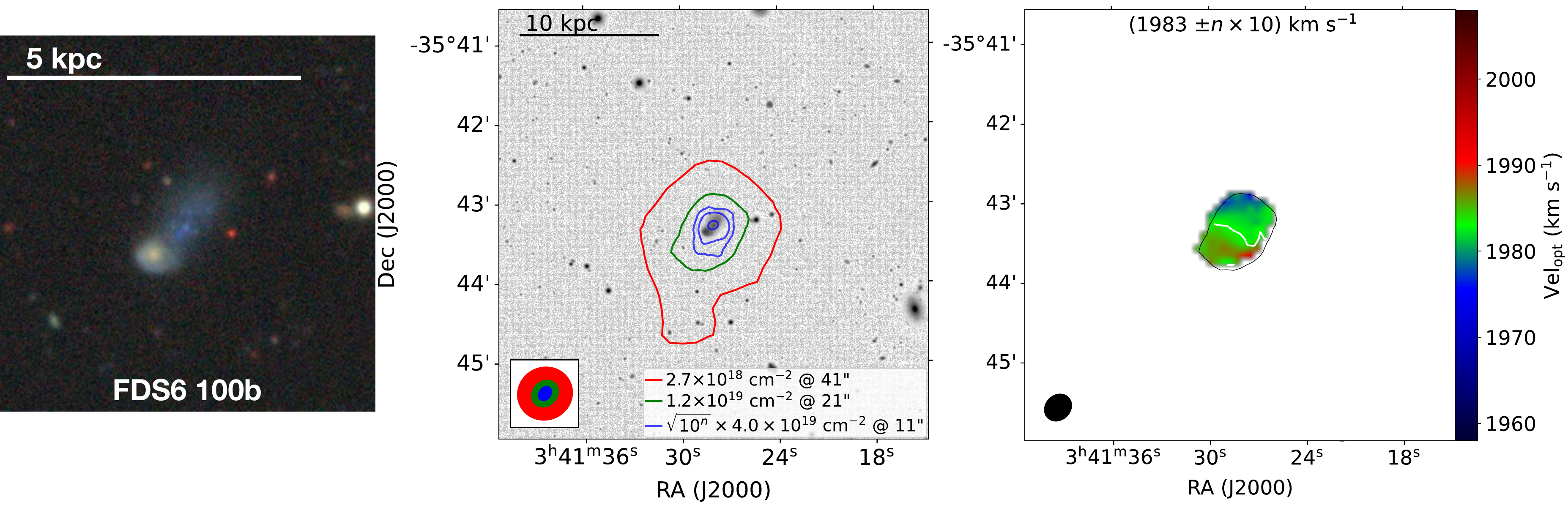}
    
  \addtocounter{figure}{-1}
   \caption{continued.}
\end{figure*}
\end{appendix}

\end{document}